\documentclass[final]{siamart1116}
\usepackage{amsmath,amsfonts,amssymb}  
\usepackage{url}  
\usepackage{epstopdf}
\usepackage{graphicx}
\graphicspath{{./Figures/}}
\usepackage{pifont}
\usepackage{pstricks, pst-node, pst-plot}
\usepackage{epsfig}
\usepackage{sidecap}
\usepackage{colortbl}
\usepackage{color}
\usepackage{booktabs}
\usepackage[inline]{enumitem}
\usepackage{tasks}

\newtheorem{prop}{Proposition}[section]

\newcommand{\bs}[1]{\boldsymbol{#1}}
\newcommand{\m}[1]{\mathbf{#1}}

\title{Transitions to Intermittent Chaos in Quorum Sensing Dynamics}

\author{
	A. Flores--P\'erez\thanks{Faculty of Engineering, UNAM, Escolar 04360, C.U., Coyoac\'an, 04510 Ciudad de M\'exico, CDMX.
	\email{afloresperez@comunidad.unam.mx}.
	Funded  by `Programa CAPSEM I+DT', Faculty of Engineering, UNAM.}
	\and
	M.~A. Gonz\'alez--Olvera\thanks{Universidad Aut\'onoma de la Ciudad de M\'exico, Dr. Garc\'ia Diego 168, Cuauht\'emoc, 06720 Ciudad de M\'exico, CDMX.
	\email{marcos.angel.gonzalez@uacm.edu.mx}. 
	Funded by project UACM CCYT2023-IMP-05 and CONAHCyT-SNII.}
	\and
	V.~F. Bre\~na--Medina\thanks{Department of Mathematics, ITAM, R\'io Hondo 1, Ciudad de M\'exico 01080, M\'exico.
	\email{victor.brena@itam.mx}.
	Funded by Asociaci\'on Mexicana de Cultura A.C.}
}

\begin{document}
\maketitle

\begin{abstract}
This study analyses the dynamical consequences of heterogeneous temporal delays within a quorum sensing-inspired (QS-inspired) system, specifically addressing the differential response kinetics of two sub-populations to signalling molecules. A nonlinear delay differential equation (DDE) model, predicated upon an activator-inhibitor framework, is formulated to represent the interspecies interactions. Key analytical techniques, including the derivation of the pseudo-characteristic polynomial and the determination of Hopf bifurcation criteria, are employed to investigate the stability properties of steady-state solutions. The analysis reveals the critical role of multiple, dissimilar delays in modulating system dynamics and inducing bifurcations. Numerical simulations, conducted in conjunction with analytical results, reveal the emergence of periodic self-sustained oscillations and intermittent chaotic behaviour. These observations emphasise the intricate relationship between temporal heterogeneity and the stability landscape of systems exhibiting  QS-inspired dynamics. This interplay highlights the capacity for temporal variations to induce complex dynamical transitions within such systems. These findings assist to the comprehension of temporal dynamics within these and related systems, and may contribute to the development of strategies aimed at modulating intercellular communication and engineering synthetic biological systems with temporal control.
\end{abstract}  



\pagestyle{myheadings}
\thispagestyle{plain}
\markboth{A.~Flores--P\'erez, M.~A. Gonz\'alez--Olvera and V.~Bre\~na--Medina}
{Transitions to Intermittent Chaos in Quorum Sensing Dynamics}

\section{Introduction}
\label{sec:intro}
Quorum sensing (QS) is a fundamental mechanism of microbial communication that was first proposed in the early 1970s to explain the bioluminescence observed in marine bacteria, particularly {\em Vibrio fischeri} \cite{nealson1970cellular}. Over the decades, QS has emerged as a sophisticated strategy that enables bacterial populations to synchronise their behaviour and enhance collective fitness. This process allows bacteria to assess their local density by monitoring the concentration of signalling molecules known as autoinducers. Once the concentration of autoinducers surpasses a critical threshold, a coordinated response is initiated, conferring advantages such as enhanced virulence \cite{lee2013cell,lyon2004peptide}, { where QS triggers the expression of virulence factors, enabling coordinated attack on a host once a sufficient bacterial community is established, as seen in {\em Pseudomonas aeruginosa} infections \cite{miller2001quorum,moura2019}. Another advantage is the colony capacity to reach biofilm formation \cite{hammer2003quorum}, a structured community of bacteria encased in a self-produced matrix, which enhances their resistance to antibiotics and host immune responses, exemplified by {\em Staphylococcus aureus} biofilms in chronic infections \cite{otto2008,parsek2003}. That is, understanding and targeting QS pathways presents promising avenues for novel antimicrobial strategies that aim to disrupt bacterial cooperation rather than directly killing them, potentially mitigating the rise of antibiotic resistance.}

The study of QS mechanisms has been benefited from diverse mathematical modelling approaches, each offering unique insights into the underlying biological dynamics. These approaches span deterministic models, which describe average system behaviours; stochastic models, which account for intrinsic fluctuations; and hybrid models that combine both perspectives. {For instance, Gorochowski~\cite{chofski2016} examines the broader scope of agent-based models in synthetic biology, including their benefits and challenges. Meanwhile, Vel\'azquez {\em et al.}~\cite{judithpv} employed a continuous-time Markov process to specifically investigate virulence in plant-pathogen interactions on leaf surfaces; their detailed model incorporated linear birth rates, logistic death-migration processes, and an autocatalytic mechanism for {\em acyl homoserine lactone} autoinducers, demonstrating an inverse relationship between QS efficiency and autoinducer diffusion.} Similarly, Frederick {\em et al.}~\cite{frederick} used a reaction-diffusion model with density-dependent diffusion coefficients to study extracellular polymeric substance (EPS) production in biofilms. They demonstrated how QS-regulated EPS production facilitates biofilms' transition from a colonisation phase to a protective state, underscoring the adaptive benefits of QS in biofilm environments.


{In particular, deterministic modelling has advanced our understanding of biological systems by incorporating the critical roles of temporal delays among other transport features not under the scope of this manuscript. Specifically, the inherent delays associated with intracellular processes, such as transcription and translation, can profoundly influence the dynamic behaviour of regulatory networks. 

The particular case of transcriptional time delays has been a subject of intense investigation. The intricate gene regulatory networks governing such mechanism, often featuring feedforward and negative feedback motifs, are particularly susceptible to the influence of these delays. Seminal works by Chen {\em et al.}~\cite{chen}, Elowitz {\em et al.}~\cite{elowitz}, and Ojalvo {\em et al.}~\cite{ojalvo} demonstrated how transcriptional delays may act as crucial parameters that affect the stability and oscillatory properties of QS circuits. Their models revealed that specific ranges of time delays can induce dynamical events such as Hopf bifurcations, leading to the emergence of sustained oscillations in the concentrations of signalling molecules and downstream effectors. This temporal orchestration has direct implications for the functional outcomes of QS, such as the precisely timed release of virulence factors in pathogenic bacteria or the coordinated formation and dispersal of biofilms.} 

{The} inclusion of time delays in QS models is crucial for accurately capturing {certain key} temporal dynamics of bacterial communication. For example, Barbarossa {\em et al.} \cite{barbarossa} developed a delay differential equations (DDE) model for {\em Pseudomonas putida} to explain experimental observations of autoinducer production. Their results supported the hypothesis of an enzyme degrading autoinducers into an inactive form, thereby introducing a feedback delay. Time delays have been shown to induce oscillatory behaviours, as demonstrated in \cite{Chen2019,Chen2020PeriodicDelay}, where the amplitude and frequency of oscillations depended on the delay length. Furthermore, bifurcation analyses of~QS models have revealed transitions between stable steady-states, periodic oscillations, chaotic behaviours ({\em e.g.}~\cite{mackey1977oscillation,guevara1983chaos,harrisvf}), and multi-stability \cite{yanchuk2009delay}. These features align with experimental observations where bacterial populations exhibit synchronised or desynchronised behaviours under varying environmental conditions \cite{goryachev}.

{Moreover, comprehension of delay-induced dynamics in QS has spurred innovation in biomedical engineering. The ability to harness QS-regulated systems for applications like controlled drug delivery in cancer therapy hinges on the precise temporal control afforded by these inherent biological delays. As demonstrated by You {\em et al.}~\cite{you2004}, cell-cell communication mechanisms, such as QS, can be engineered to achieve programmed population control, highlighting the potential for temporally regulated biological processes. Building upon this, Tamsir {\em et al.}~\cite{tamsir2011} showcased robust multicellular computing using synthetic biology, where QS enables coordinated behaviour within a cell population, demonstrating the capacity for controlled actions in biological systems. The inherent time lags in QS responses are crucial for achieving this level of control.}

Incorporating these temporal factors into QS models not only enriches our understanding of microbial communication but also strengthen the development of practical applications. Manipulating QS systems can lead to novel strategies for microbial control and synthetic biology, with potential implications for environmental management and medical therapeutics \cite{fergola2006allelopathic,fergola2008influence,Zhang20082834,Zhonghua2009201}. For instance, in \cite{Chen2019}, a single delay introduced into a QS model led to sustained oscillations in signalling molecule concentrations, highlighting the pivotal role of temporal factors in QS regulation. Moreover, Chen {\em et al.} \cite{chen} analysed the effects of two distinct delays: one representing the time required for autoinducers to diffuse and another capturing the lag between signal detection and target gene expression. Their results demonstrated the importance of these delays in driving oscillatory behaviours and determining system stability.


{In bacterial QS, certain receptor proteins perceive autoinducers, initiating intricate gene expression programmes that govern collective behaviours. This often results in the emergence of functionally distinct sub-populations: a motile or active fraction, where genes related to collective activities are significantly up-regulated, and a static or reservoir fraction, where these genes are correspondingly down-regulated. The motile subpopulation actively contributes to processes like biofilm formation, virulence factor secretion, or bioluminescence, while the static subpopulation may serve as a reserve, potentially contributing to population survival under stress or acting as a source for future active cells. However, the transitions between these dynamic states are not immediate consequences of autoinducer detection. Instead, inherent time delays, arising from fundamental biochemical processes such as the transcription of genes, the translation of messenger RNA into proteins, and the maturation or activation of these proteins, introduce a temporal dimension to QS responses~\cite{lewis2003autoinhibition,monk2003oscillatory,zhang2016oscillatory,zhang2017oscillatory,Mukherjee}. These delays, which encapsulate the time scales of signal production, extracellular diffusion, receptor binding, intracellular signalling cascades, and the eventual transcriptional and translational responses, critically modulate the timing, synchronisation, and robustness of QS-mediated collective behaviours~\cite{Mukherjee}. Even though there is intense research of time delay systems occurring in other fields as reported in~\cite{arbi2022a,arbi2022b}, for instance, we focus on understanding these delay-dependent dynamics as crucial for deciphering the complex temporal organisation particularly within bacterial communities and for developing strategies to manipulate QS-controlled processes.}

{In the present manuscript, we analyse a nonlinear DDE-QS system wherein motile and static bacterial sub-populations exhibit differential responses to autoinducers due to distinct response times. This is characterised by the introduction of two independent delay parameters, $\tau_1 > 0$ and $\tau_2 > 0$, corresponding to the motile and static sub-populations, respectively. A local stability analysis is conducted by varying these delay parameters, identifying diverse bifurcations and the conditions under which they arise. These findings contribute to the understanding of how time delays shape the dynamics of QS systems, particularly concerning bifurcations and stability transitions. Furthermore, the emergence of chaotic dynamics from such transitions holds the potential to significantly clarify crucial bacterial population behaviour, potentially enabling the identification of key conditions leading to disruptive QS, among other phenomena. This parallels how chaos theory has illuminated key features in other systems, as explored in~\cite{cicek,ozcelik}.

The theoretical framework for deterministic chaos in biological systems characterised by nonlinear feedback and time delays was established by the seminal works of Mackey and Glass~\cite{mackey1977oscillation} and Guevara {\em et al.}~\cite{guevara1983chaos}. These investigations demonstrated that complex temporal dynamics could arise endogenously from such system architectures, a concept further explored in the context of biological oscillators~\cite{may1974}.

Mathematical modelling, as evidenced in studies demonstrating bifurcations and chaos in synthetic QS systems and analysing chaotic behaviours in delayed reaction-diffusion QS models, reveals the potential for chaotic regimes within QS networks under specific parametric conditions. This is consistent with the broader understanding that nonlinearities and time delays inherent in QS mechanisms can lead to complex dynamics~\cite{hellen2018,karkaria2022}. Complementing these findings, Harris {\em et al.}~\cite{harrisvf} provide a mechanistic understanding of these complex behaviours by detailing the role of local and global bifurcations in facilitating transitions between stable states and potentially chaotic invariant sets. Their analysis of bifurcation phenomena elucidates how parameter variations within QS systems can lead to dynamic instability and complex outputs.

The broader functional implications of such dynamic complexity for bacterial adaptation and resilience are reflected in the investigation of hyperchaos in coupled QS systems. Stankevich and Volkov~\cite{hyperchaos} suggest that interactions between bacterial units or QS circuits can amplify dynamical complexity. The potential for chaotic dynamics to offer a form of robustness or adaptability for bacterial communities, allowing for a wider exploration of behavioural states, aligns with the theoretical possibilities discussed, for instance, in~\cite{Chen2020PeriodicDelay,hellen2018,karkaria2022}.}

The manuscript is structured as follows: Section~\ref{sec:lind} introduces the nonlinear time-delayed QS system under investigation, detailing the model's assumptions and highlighting the activator-inhibitor framework employed to represent the dynamic interplay between motile and static bacterial sub-populations. Section~\ref{sec:second} rigorously analyses the existence and local stability properties of the system's steady-states, deriving the pseudo-characteristic polynomial and formulating conditions for the emergence of purely imaginary eigenvalues, thereby facilitating the analysis of Hopf bifurcations. Section~\ref{sec:third} presents the results of extensive numerical simulations, exploring the system's dynamical behaviour across a range of parameter values and illustrating the profound influence of varying delay parameters on the system's dynamics. A detailed analysis of the observed intermittent chaotic behaviour, including a discussion of its characteristics and underlying mechanisms, is presented in Section~\ref{sec:four}. Finally, Section~\ref{sec:five} concludes the manuscript with a comprehensive discussion of the key findings, emphasising their implications for advancing our understanding of QS-inspired systems and, in particular, the emergence of complex dynamical behaviour in the presence of multiple delays.


\section{Nonlinear time delayed Quorum Sensing system }
\label{sec:lind}
The present work investigates the dynamics of QS, emphasising the critical role of local interactions mediated by autoinducer concentrations. Autoinducers are signalling molecules produced by both motile and static bacteria, following the law of mass action, whose production rates increase with bacterial population density. The growth of motile bacteria is intrinsically driven as well as is inhibited by the presence of static bacteria, with long-term behaviour regulated by environmental saturation constraints.

The bacterial population dynamics in this context can be modelled using an activator-inhibitor framework analogous to the Gierer--Meinhardt system. Here, motile bacteria function as activators, driving QS processes, while static bacteria act as inhibitors, imposing regulatory constraints. To reduce the complexity of the parameter space and address scenarios where static bacteria may be negligible, rescaling transformations and desingularisation techniques are employed. The resulting QS model is governed by the following system of equations, as described in~\cite{harrisvf}:
\begin{subequations}\label{eq:qs}
\begin{flalign}
\dot{w} & = dv(u+v) - cwv, \label{qsa} \\[1ex]
\dot{v} & = wvu^2 + a\varepsilon v - v^2, \label{qsb} \\[1ex]
\dot{u} & = \frac{wu^2}{1+Ku^2} + av - buv, \label{qsc}
\end{flalign}
\end{subequations}
where $w$, $v$, and $u$ represent the concentrations of autoinducers, static bacteria, and motile bacteria sub-populations, respectively. The parameters are defined as follows:  
\begin{enumerate*}[label=(\roman*)]
\item $K$ is the saturation parameter, modulating the regulatory effects of autoinducer concentrations;  
\item $a$ and $\varepsilon a$ denote the production rates of autoinducers and the bacterial population, respectively;  
\item $b$ is the decay rate of the bacterial population; and  
\item $c$ and $d$ describe the decay and production rates of autoinducers.  
\end{enumerate*}

The system in~\eqref{eq:qs} encompasses the dynamic interplay between motile and static bacteria populations, governed by their interactions with autoinducers. Specifically:  
\begin{enumerate*}[label=(\roman*)]
\item the rate of change of the autoinducer concentration depends on the production and decay rates driven by interactions between motile and static bacteria;
\item the static bacteria concentration evolves according to its interactions with autoinducers and its own production-decay dynamics;  
\item the motile bacteria concentration is influenced by its interactions with autoinducers and static bacteria, along with intrinsic production and decay rates.  
\end{enumerate*}

This study advances a rigorous framework, predicated upon an activator-inhibitor paradigm, for the comprehensive analysis of QS regulatory mechanisms and localised interaction dynamics. By explicitly formulating the dependence of bacterial population dynamics on cell densities and autoinducer-mediated feedback loops, the framework captures the salient features of QS behaviour. Notably, the application of rescaling transformations enhances the model's applicability, particularly in scenarios characterised by the absence of static bacterial populations.

Beyond its methodological contributions, this work provides novel insights into the nuanced dynamics of microbial communication within spatially constrained environments. The model elucidates the synergistic interplay between localised interactions and intrinsic regulatory mechanisms in shaping emergent bacterial population dynamics, offering critical perspectives on the complex, density-dependent behaviours characteristic of QS systems. These findings establish a robust theoretical foundation for the systematic investigation of QS dynamics across diverse environmental and biological contexts, facilitating advancements in fields such as microbial ecology, systems biology, and biomedical engineering. Specifically, the rescaling transformations further enhance the model's applicability, particularly in cases where static bacteria may be absent.

{Given that bacterial responses to autoinducer-mediated stimuli are not immediate, it is plausible that the ensuing cellular activities are contingent upon intrinsic receptor characteristics within distinct sub-populations or on physiological processes enacted at preceding times.  While acknowledging the biological limitations of relying solely on constant delays, their initial and selective use in QS models is supported by some pragmatic advantages. Firstly, the mathematical simplicity and analytical tractability of DDEs with constant delays provide a robust framework for initial investigations into system dynamics, which allows to readily explore stability landscapes, identify potential bifurcations, and characterise oscillatory regimes, offering qualitative insights into the fundamental impact of temporal lags on QS behaviour. Secondly, employing constant delays can serve as a crucial first-order approximation of the system's temporal characteristics. Upon capturing the non-instantaneous nature of signal transduction and gene expression, this simplified model provides a foundational understanding of how time scales influence QS responses, particularly when detailed kinetic data for more complex delay formulations are unavailable. Finally, in specific contexts when the modelling focus is on emergent macroscopic population-level phenomena as is in the present case, constant delays can offer a reasonably accurate representation without the added complexities of distributed or state-dependent delays. Consequently, while not offering a comprehensive depiction of biological reality, constant delays can serve as an entry point for exploring the temporal intricacies of QS.}

To account for these temporal effects {as well as the delays features already described in Section~\ref{sec:intro}}, the original system~\eqref{eq:qs} is extended by incorporating delayed autoinducer concentrations into the growth rates. Specifically, let $w_1 = w(t-\tau_1)$ and $w_2 = w(t-\tau_2)$, where $\tau_1 > 0$ and $\tau_2 > 0$ represent the delay times associated with the non-instantaneous responses of motile and static bacteria, respectively. The modified QS system is then given by:
\begin{subequations}\label{eq:qsd}
\begin{flalign}
\dot{w} &= dv(u+v) - cwv, \label{qsad} \\[1ex]
\dot{v} &= w_1 vu^2 + a\varepsilon v - v^2, \label{qsbd} \\[1ex]
\dot{u} &= \frac{w_2 u^2}{1+K u^2} + av - buv, \label{qscd}
\end{flalign}
\end{subequations}
where the parameters retain their original definitions, as described earlier, and their distinguished values are as in Table~\ref{table:Ap1}.

Notice that the inclusion of delays $\tau_1$ and $\tau_2$ introduces additional complexity to the system dynamics, as these delays reflect the temporal lag between autoinducer production and its regulatory effects on bacterial populations. {The delayed terms $w_1 = w(t - \tau_1)$ and $w_2 = w(t - \tau_2)$ capture} the influence of past autoinducer concentrations on the current growth rates of static and motile bacteria, respectively. Our analysis focuses on examining the dynamic impact of these delays, particularly how they influence stability and potential bifurcation phenomena. 

\begin{table}[t]
\centering
  \begin{tabular}{ |c|c|c|c|c|c|c|c|}
    \hline
    $a$ &  $b$ & $\varepsilon$ & $c$ & $d$ & $K$ & $\tau_1$ & $\tau_2$ \\ \hline
    $0.0428$ & $[1.2,1.5]$ & $0.1$ & $0.16$ & $0.0522$ & $0.0435$ & $[0,500]$ & ${[0,1000]}$ \\ \hline
  \end{tabular}
  \caption{Parameter values set { as given in~\cite{harrisvf}, and a wide range for delay parameters $\tau_{1,2}$}.}
  \label{table:Ap1}
\end{table}

\section{Steady-states and local stability.}
\label{sec:second}

To initiate our analysis, we first establish the existence and positivity of steady-states under the parameter values specified in Table~\ref{table:Ap1}. Subsequently, we proceed to examine the local stability properties of these steady-states to gain insights into the system's dynamic behaviour.

\subsection{Positive steady-states.}
\label{sec:steady-states}

Biologically meaningful steady-states of system~\eqref{eq:qsd} correspond to constant solutions that lie within the first octant. {Necessarily, such a solutions are determined by the condition where the time derivatives on the left-hand side of either system~\eqref{eq:qs} or system~\eqref{eq:qsd} vanish. That is, these steady-states are thus} determined by solving the following algebraic system: 
	\begin{subequations}\label{eq:equilibrium}
	\begin{flalign}
		&\left[(u+v)d - cw\right]v = 0\,, \label{qsae}\\[1ex]
		&\left(wu^2 + a\varepsilon  - v\right)v = 0\,, \label{qsbe}\\[1ex]
		&wu^2 + \left(a - bu\right)\left(1 + Ku^2\right)v = 0\,, \label{qsce}
	\end{flalign}
	\end{subequations}
where $w_1 = w_2 = w$, since steady-states are temporally invariant. 

The null steady-state $(w, v, u) = (0, 0, 0)$ corresponds to the absence of bacterial populations and autoinducer concentration. For non-trivial steady-states, equation~\eqref{qsae} provides the following expression for $w$: 

\begin{subequations}\label{eq:wvsteady}
\begin{gather}\label{wdes}
    w = \frac{d(u + v)}{c} > 0\,,
\end{gather}
which can be substituted into~\eqref{qsbe}, yielding the expression for $v$: 
\begin{gather}\label{vdes}
    v = \frac{du^3 + a\varepsilon c}{c - du^2}\,,
\end{gather}
\end{subequations}
provided that $|u| < \sqrt{c/d}$ is satisfied to ensure $v > 0$. Substituting~\eqref{wdes} and~\eqref{vdes} into~\eqref{qsce} reduces the system to solving $f_{eq}(u) = 0$, where $f_{eq}(u)$ is a monic polynomial of degree six, given by

\begin{flalign}\label{upol}
    f_{eq}(u) := &\: u^6 - \frac{a}{b}u^5 + \frac{1}{K}u^4 - \frac{ad + d - ba\varepsilon cK}{bdK}u^3 \\[1ex]
    & - \frac{da\varepsilon + a^2\varepsilon cK}{bdK}u^2 + \frac{a\varepsilon c}{dK}u - \frac{a^2\varepsilon c}{bdK}\,. \nonumber
\end{flalign}
Since all coefficients of $f_{eq}(u)$ are real, the Fundamental Theorem of Algebra guarantees six roots in total, which may include real and complex ones. The number of positive, negative, and complex roots is constrained by Descartes' rule of signs. That is, based on the five sign changes in the polynomial’s coefficients, provided $(a+1)d>ba\varepsilon cK$, three possible configurations arise: 
\begin{enumerate*}[label=(\roman*)]
    \item five positive roots and one negative root;
    \item three positive roots, one negative root, and two complex conjugate roots;
    \item one positive root, one negative root, and four complex conjugate roots.
\end{enumerate*}

To refine this further, the Sturm Theorem 
can be applied to determine the number of distinct real roots of~\eqref{upol} within the interval $|u| < \sqrt{c/d}$. 
Using the parameter values given in Table~\ref{table:Ap1} with $b=1.4$, the signs of the Sturm sequence are calculated at the ends of the interval of consideration. 
{Consequently, the number of real roots within the interval is four.} 
Combining this result with Descartes' rule of signs confirms that the roots of $f_{eq}(u)$ comprise three positive, one negative, and two complex conjugate roots. In consequence, there are three steady-states lying within the first octant. { This yields to the following proposition 

\begin{prop}
For system~\eqref{eq:equilibrium}, its associated monic polynomial of degree six, defined in \eqref{upol}, possesses three positive real roots, one negative real root, and two complex conjugate roots if and only if $(a+1)d > ba\varepsilon cK$. Under this condition, and within the interval $|u| < \sqrt{c/d}$, all positive steady-states of system~\eqref{eq:qsd} are determined.
\end{prop}
}

\subsection{Time delayed QS local stability.}

{Results presented in~\cite{harrisvf} reveal oscillatory dynamics within a non-delayed framework in the context of this model, but significantly, they also uncover chaotic dynamics explained by homoclinic Shilnikov bifurcations (see, for instance,~\cite{paguirre}), which arise from a primary Hopf bifurcation. These bifurcation findings are shown to be dependent on external rather than internal parameters.

In contrast, our approach here focuses on disclosing the consequences of slowly varying internal delay parameters. Unlike the simplification of a single, constant delay---which can introduce temporal effects and oscillations absent in non-delayed systems---we acknowledge that real biological systems often exhibit a distribution of delays. Consequently, a single constant value inherently fails to capture the richer information encoded within these more realistic delay distributions. Such a simplification can prevent the model from reproducing more complex dynamics, including intricate oscillatory patterns, bifurcations triggered by delay magnitude, or multi-stability, which can arise when a range of delays is considered. Thus, our initial exploration using a single delay (not shown) proved inadequate to sufficiently portray the complex temporal regulation within the QS system under study.

That is, while Harris {\em et al.} in~\cite{harrisvf} provide a detailed analysis of Hopf bifurcations in the non-delayed QS model version, the general principles are fundamental to understanding the more complex dynamics of DDEs. The presence of delays in QS systems adds a layer of complexity, making the study of bifurcations crucial for deciphering and potentially controlling bacterial behaviour. Let us recall the crucial points of Hopf bifurcation~\cite{michiels2007stability,nguimdo,suzuki2016periodic,Wernecke_2019}:
\begin{enumerate*}[label=(\roman*)]
    \item As can be seen further, the stability of a steady-state is determined by a characteristic equation that is transcendental, unlike the polynomial equation in ordinary differential equations (ODEs). This transcendental equation has infinitely many roots (eigenvalues), making the analysis more intricate.
    \item Similar to ODEs, a Hopf bifurcation in DDEs occurs when a pair of complex conjugate eigenvalues of the linearised system cross the imaginary axis as a parameter slowly varies, which is known as a codimension-one bifurcation. However, in DDEs, this crossing can happen multiple times for different eigenvalue pairs as the delay is increased, leading to multiple secondary bifurcations and complex dynamics.
    \item The delay itself is a critical bifurcation parameter in DDEs. Varying the delay can induce oscillations and instability even if the steady-state for the corresponding ODE system is stable, as well as affecting the frequency of the emerging oscillations, as it is determined by the imaginary part of the eigenvalues crossing the imaginary axis. This is a key difference from ODEs, where the delay is absent.
    \item The stability of the bifurcating periodic solutions is determined by the direction of the eigenvalue crossing and the non-linear terms of the system. This can be analysed using normal form theory and the computation of Lyapunov coefficients, as shown, for instance, in~\cite{harrisvf} for the non-delayed case. We here propose an alternative approach.
\end{enumerate*}

Hence, to analyse the local stability properties of the system, let us denote the positive steady-states discovered in Section~\ref{sec:steady-states} as $E_1,E_2,E_3\in\mathbb{R}^3$.}

{
We consider the nonlinear DDE system defined by $\dot{\mathbf{x}}=\mathbf{F}(\mathbf{x},\mathbf{x}_{\tau_1},\mathbf{x}_{\tau_2})$. Here, the non-delayed state vector is $\mathbf{x}=(w,v,u)\in\mathbb{R}^3_+$, and $\mathbf{x}_{\tau_1}=(w_1,0,0)$ and $\mathbf{x}_{\tau_2}=(w_2,0,0)$ denote the two time-delayed states, associated with positive time delays $\tau_1,\tau_2>0$. The function $\mathbf{F}=(f_1,f_2,f_3)$ is a differentiable vector-field, and its components correspond to the right-hand side of system~\eqref{eq:qsd}. Analogous to the non-delayed case, a linearisation approach is applied to this retarded nonlinear system. This technique relies on constructing local estimations of the nonlinear states in the vicinity of steady-states. Let 
\begin{gather*}
	\left(\mathbf{J}_0\right)_{i,j}=\dfrac{\partial f_i}{\partial x_j}\,, \quad i,j=1,2,3\,,
\end{gather*}
denote the matrix of system partial derivatives with respect to the non-delayed states. Similarly, let matrices 
\begin{gather*}
	(\mathbf{J}_{\tau_k})_{i,j}=\dfrac{\partial f_i}{\partial x_{\tau_{kj}}}\,, \quad  i,j=1,2,3 \quad \textrm{and} \quad k=1,2\,,
\end{gather*}
account for the local contributions of the non-immediate variables arising from the time delays $\tau_1,\tau_2>0$. The {\em ansatz} of an exponential solution for the linearised system necessitates a nontrivial evolution near steady-states only when there exist complex values $\lambda$ such that the so-called \textit{pseudo-characteristic polynomial} satisfies (see~\cite{michiels2007stability}):
} 
\begin{subequations}\label{detd}
\begin{gather}
    \det(\mathbf{J}_0 + \exp(-\lambda \tau_1) \mathbf{J}_{\tau_1} + \exp(-\lambda \tau_2) \mathbf{J}_{\tau_2} - \lambda \mathbf{I}_{3 \times 3}) = 0\,, \label{detda}
\end{gather}
{
where $\mathbf{I}_{3 \times 3}$ is the identity matrix and exponential factors arise given the dependence on the history of variables within past intervals defined by~$\tau_1$ and~$\tau_2$. 
 
 In the particular case of QS time delayed system \eqref{eq:qsd}, matrices $\mathbf{J}_0$, $\mathbf{J}_{\tau_1}$, and $\mathbf{J}_{\tau_2}$ are given by 
 } 
\begin{gather}
    \mathbf{J}_0 = 
    \begin{pmatrix}\label{detdb}
        -cv & du + 2dv - cw & dv \\ 
        0 & w_1 u^2 + a\varepsilon - 2v & 2w_1 vu \\ 
        0 & a - bu & \frac{2uw_2}{(1 + Ku^2)^2} - bv
    \end{pmatrix}\,, \\[1.5ex]
    \mathbf{J}_{\tau_1} = 
    \begin{pmatrix}\label{detdc}
        0 & 0 & 0 \\ 
        vu^2 & 0 & 0 \\ 
        0 & 0 & 0
    \end{pmatrix}\,, \quad 
    \mathbf{J}_{\tau_2} = 
    \begin{pmatrix}
        0 & 0 & 0 \\ 
        0 & 0 & 0 \\ 
        \frac{u^2}{1 + Ku^2} & 0 & 0
    \end{pmatrix}\,.
\end{gather}
\end{subequations}
The matrix $\mathbf{J}_0$ corresponds to the Jacobian evaluated at the steady-state in the absence of delays, while $\mathbf{J}_{\tau_1}$ and $\mathbf{J}_{\tau_2}$ capture the contributions from delayed interactions {associated with $w_1$ and $w_2$}, respectively. These delay terms reflect the biological reality that bacterial responses to autoinducers may not be instantaneous, depending on the activity of each subpopulation or autoinducers response characteristics.

The determinant condition in~\eqref{detd} incorporates the interplay of delays in the system's dynamics, where the $\tau_{1,2}$-dependent exponential terms introduce non-trivial dependencies on the delay parameters. These factors fundamentally affect the stability of the steady-states, as they influence the eigenvalue spectrum of the pseudo-characteristic equation.

Notice that the structure of the matrices $\mathbf{J}_0$, $\mathbf{J}_{\tau_1}$, and $\mathbf{J}_{\tau_2}$ highlights the coupling between up- and down-regulated bacteria, and autoinducer concentrations. In other words,  
\begin{enumerate*}[label=(\roman*)]
\item the matrix~\eqref{detdb} encapsulates the direct interactions and self-regulation of each variable;
\item the left-hand side delay matrix in~\eqref{detdc} accounts for the delayed contribution of motile bacteria to the growth of static bacteria via autoinducer-mediated interactions;
\item similarly, right-hand side delay matrix in~\eqref{detdc} reflects the delayed influence of motile bacteria on their own growth due to the saturation effects regulated by autoinducer concentration.  
\end{enumerate*}

As the pseudo-characteristic polynomial~\eqref{detd} defines a transcendental equation, it inherently possesses an infinite number of roots due to the presence of exponential terms. These roots, corresponding to the eigenvalues of the linearised system, determine the growth rates and oscillatory behaviours of the dynamics. However, such linearisation at any steady-state is characterised by at most a finite number of eigenvalues with positive real parts, which correspond to unstable modes. The remaining eigenvalues typically have negative real parts or tend asymptotically to $-\infty$. As a consequence, despite the infinite-dimensional nature of system~\eqref{eq:qsd}, which arises from their dependence on the history of the solution over the delays interval, bifurcation theory reveals that their nonlinear dynamical events are qualitatively analogous to those of ordinary differential equations; see, for instance,~\cite{lakshmanan2011dynamics,Calleja2017ResonanceDelays}. 

{Observe that, upon substituting~\eqref{detdb} and~\eqref{detdc} into~\eqref{detda}, we get the pseudo-polynomial   }
\begin{subequations}\label{quasy}
\begin{flalign}
q(\lambda,\boldsymbol{\tau},\mathbf{p})=&\:p_3(\lambda,\mathbf{p})+p_2(\lambda,\mathbf{p})\exp(-\lambda\tau_2)+p_1(\lambda,\mathbf{p})\exp(-\lambda\tau_1)\,,\label{quasyp}
\end{flalign}
where
\begin{flalign}
p_3(\lambda,\boldsymbol{\tau},\mathbf{p})=&\: (-cv-\lambda) \bigg[\big(a\varepsilon-\lambda+u^2 {w_1}-2v\big) \bigg(-bv-\lambda+\frac{2u {w_2}}{\big(K u^2+1\big)^2}\bigg)\nonumber\\
&\quad -2u v {w_1}(a-b u)\bigg]\,,\label{p3}\\
p_2(\lambda,\boldsymbol{\tau},\mathbf{p})=&\:\frac{u^2 v}{K u^2+1} \big(-ad\varepsilon-2c u w {w_1}-d u^2 {w_1}+4d u v {w_1}+2d v+d \lambda\big),\label{p2}\\
p_1(\lambda,\boldsymbol{\tau},\mathbf{p})=&\: -u^2 v (-cw+du+2dv) \bigg(-bv-\lambda+\frac{2u {w_2}}{\big(K u^2+1\big)^2}\bigg)\nonumber\\
&\quad +d u^2 v^2 (a-b u)\,;\label{p1}    
\end{flalign}
\end{subequations}
here, $\boldsymbol{\tau} = (\tau_1, \tau_2)$ represents the delay vector, while $\mathbf{p} = (a, b, c, d, \varepsilon, K)$ encapsulates the system parameter vector. 

\subsubsection{Existence of purely imaginary eigenvalues.}

We begin our analysis by considering the non-delayed case in~\eqref{quasyp}. In so doing, it is reduced to the polynomial determined by~\eqref{quasy} with $\tau_1=\tau_2=0$. For the parameter values given  in Table~\ref{table:Ap1}, the positive steady-states $E_1$, $E_2$, and $E_3$ correspond to hyperbolic points. Specifically, $E_2$ is a saddle point, while $E_3$ corresponds to a stable node. In particular, the steady-state~$E_1$ exhibits a saddle-type instability as is characterised by the presence of one real negative eigenvalue and a pair of complex conjugate eigenvalues with positive real parts. Furthermore, the number of positive steady-states and their corresponding stability properties align precisely with the results reported in~\cite{harrisvf}, which states the robustness and validity of our theoretical framework.


For sufficiently small positive values of the delays, $0<\tau_1,\tau_2\ll1$, the number of eigenvalues $\lambda$ satisfying $q(\lambda,\bs{\tau},\m{p})=0$ not only becomes infinite, but also the real parts of these eigenvalues drift: they shift to the right when~$\Re{(\lambda)}$ is negative and to the left when $\Re{(\lambda)}$ is positive. This phenomenon suggests a transition in stability, which is intrinsically linked to the emergence of purely imaginary eigenvalues; see, for a detailed discussion,~\cite{michiels2007stability}.
 
To determine purely imaginary roots of \eqref{quasyp}, let us set $\lambda = i\omega$ for $\omega > 0$, yielding  
\begin{equation}\label{quasyp2}
 p_3(i\omega, \m{p}) + p_2(i\omega, \m{p}) e^{-i\omega\tau_2} + p_1(i\omega, \m{p}) e^{-i\omega\tau_1}=0\,.
\end{equation}  
First, observe that \eqref{p3} has no zeros on the imaginary axis at the steady-states $E_1, E_2, E_3$ for the parameter values given in Table~\ref{table:Ap1}. Consequently, defining $ q_i(i\omega) = p_i(i\omega, \m{p})/p_3(i\omega, \m{p}) $ for $ i = 1,2 $, where we have omitted explicit dependence on $\m{p}$ to ease notation, the pseudo-polynomial in~\eqref{quasyp2} can be rewritten as  
\begin{subequations}\label{QuasiSum}
\begin{gather}\label{quasySum1}
	1+q_1(i\omega) e^{-i\omega\tau_1}+q_2(i\omega) e^{-i\omega\tau_2} = 0\,,
\end{gather}  
which leads to  $|q_1(i\omega) e^{- i\omega\tau_1} + q_2(i\omega) e^{- i\omega\tau_2}| = 1$.
Notice that, upon introducing the definitions  $A(\omega) = \operatorname{Re}(\overline{q_1(i\omega)} q_2(i\omega))$,  $B(\omega) = -\operatorname{Im}(\overline{q_1(i\omega)} q_2(i\omega))$ and $h = \tau_1 - \tau_2$,
it can be recasted as  
\begin{gather} \label{quasyOmega}
    |q_1(i\omega)|^2 + |q_2(i\omega)|^2 + 2R(\omega) \cos(\omega h - \alpha(\omega)) = 1\,,
\end{gather}  
\end{subequations}
where  $R(\omega) = \sqrt{A^2(\omega) + B^2(\omega)}$ and $\tan \alpha(\omega) = B(\omega)/A(\omega)$.

From equation \eqref{quasyOmega}, the existence of roots of $ q(i\omega, \boldsymbol{\tau}, \mathbf{p}) $ is ensured if  the following conditions hold uniquely and simultaneously:  
\begin{subequations}\label{quasyDef}
\begin{gather}
\left| 1 - |q_1(i\omega)|^2 - |q_2(i\omega)|^2 \right| \leq 2R(\omega)\,,\label{Hopf1a}\\[1ex]
\xi(\omega) = \cos(\omega h - \alpha(\omega))\,,\label{Hopf2a}
\end{gather}
\end{subequations}  
where  
\begin{equation*}
\xi(\omega) = \frac{1 - |q_1(i\omega)|^2 - |q_2(i\omega)|^2}{2R(\omega)}\,.
\end{equation*}  
{ As a direct consequence of this result, we state:
\begin{prop}
The pseudo-polynomial defined by equation~\eqref{quasy} possesses purely imaginary roots if and only if conditions~\eqref{quasyDef} are met.
\end{prop}
}
\begin{figure}[t!]
    \centering
    \includegraphics[scale=0.3]{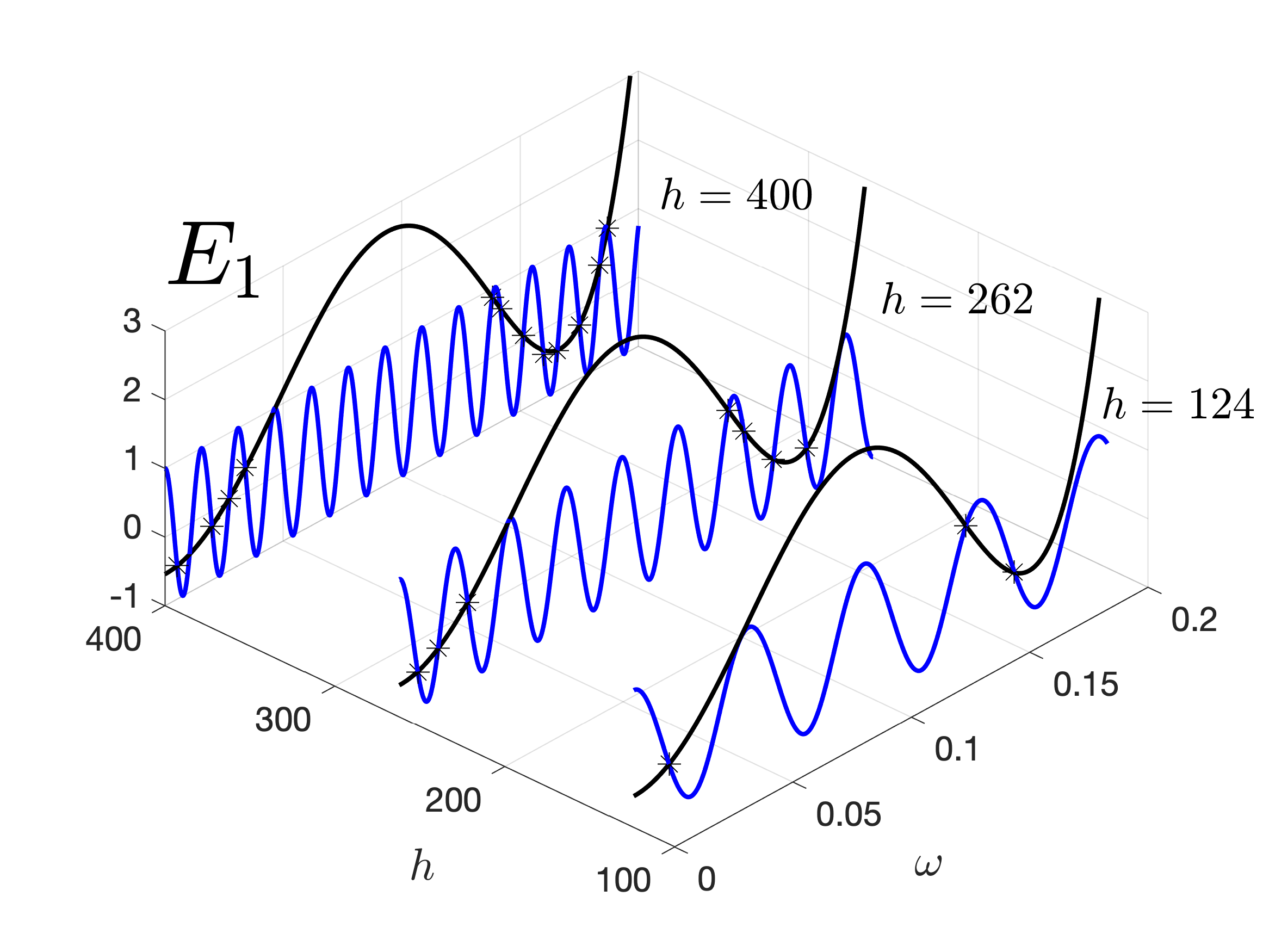} \  \includegraphics[scale=0.3]{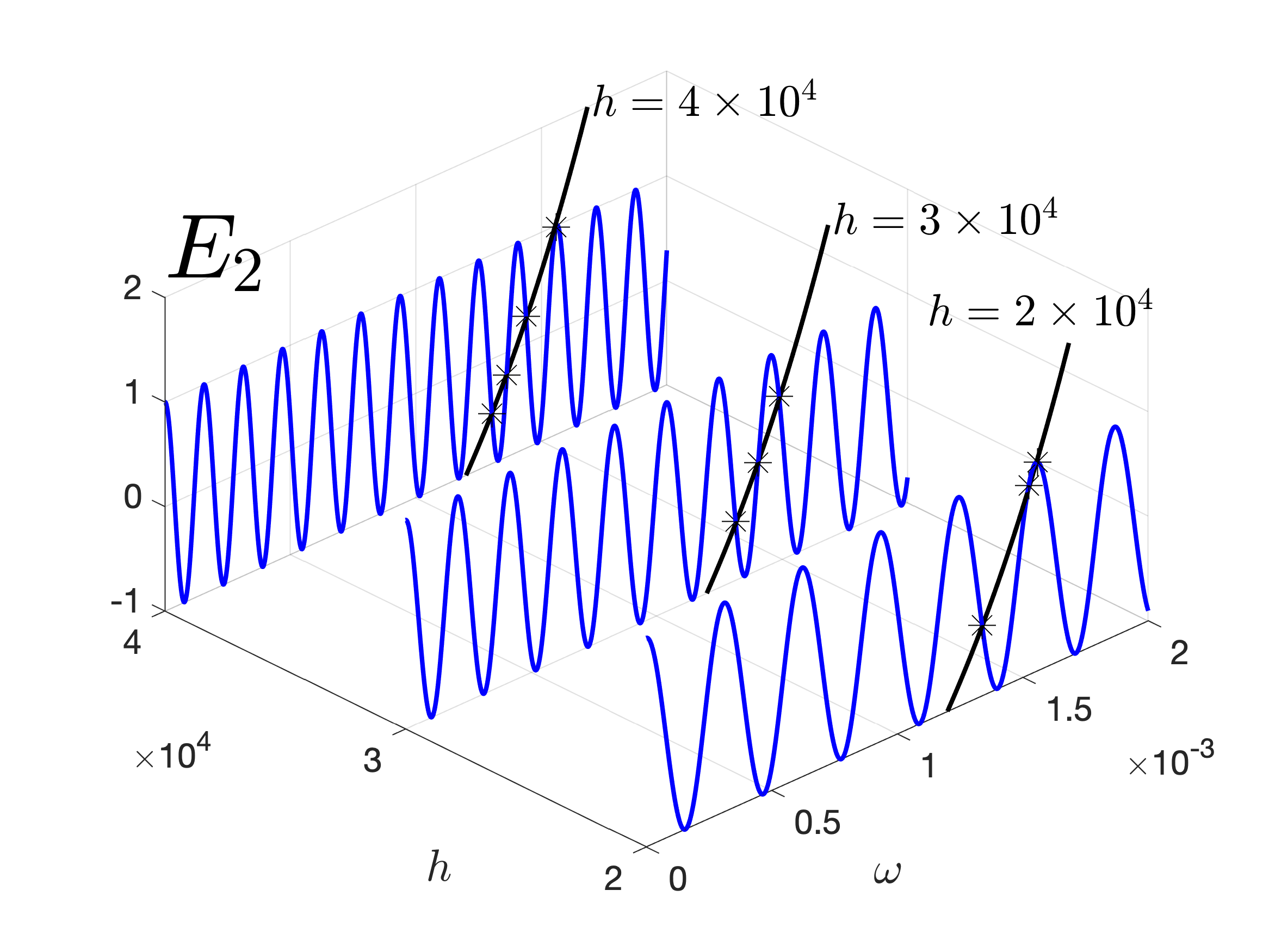}\\
    $(a)$ \hspace{5.5cm} $(b)$ 
    \caption{Existence of purely imaginary eigenvalues for the pseudo-polynomial~\eqref{quasy}. Such solutions are guaranteed for $\omega>0$ values that satisfy \eqref{quasyDef}; that is, where functions $\xi(\omega)$ (black curve) and $\cos(\omega h-\alpha(\omega))$ with fixed $h$ (blue curve) intersect. Cases for equilibria $E_1$ and $E_2$ are considered separately. (a)~Analysis of intersections at $E_1$ for $h$ values of $124, 262, 400$. (b)~Similar solution visualisations for $E_2$ with $h=2\times10^{4}, 3\times10^{4}. 4\times10^{4}$.  Parameter values as in Table~\ref{table:Ap1}. 
    }
    \label{fig:E1Study}
\end{figure}

Condition \eqref{Hopf1a} ensures that $ |\xi(\omega)| \leq 1 $, which constitutes a necessary criterion for the solvability of equation \eqref{Hopf2a}. Consequently, the roots of~\eqref{quasyp2} correspond to values of $ h \in \mathbb{R} $ and $ \omega > 0 $ at which the functions $ \xi(\omega) $ and $ \cos(\omega h - \alpha(\omega)) $ intersect.  This behaviour is depicted in Figure~\ref{fig:E1Study}(a), which illustrates the intersection points for the steady-state $ E_1 $ at three distinct values of $ h $. A similar pattern is observed for the steady-state $ E_2 $, as is shown in Figure~\ref{fig:E1Study}(b). Notably, the orders of magnitude of $ h $ at which these intersections occur differ significantly between the two steady-states: for $ E_1 $, intersections arise at values of $ h $ on the order of $ 10^2 $, whereas for $ E_2 $, they occur at values two orders of magnitude greater. In both cases, the frequency of oscillations in $ \cos(\omega h - \alpha(\omega)) $ increases proportionally with $ h $, facilitating the number of intersections, which suggests that the emergence of purely imaginary roots is strongly influenced by the introduction of two distinct delay terms, particularly when their magnitudes differ considerably. {This suggests a greater robustness in the stability of equilibrium $E_2$ compared to~$E_1$.} On the other hand, the steady-state $ E_3 $ is evaluated in the functions $ \xi(\omega) $ and $ \cos(\cdot) $. In this case, it is disclosed that $ |\xi(\omega)| > 1 $ for all $ \omega \geq 0 $. Consequently, condition~\eqref{Hopf1a} is hence not satisfied, preventing the existence of intersection solutions for~\eqref{Hopf2a}, which results in confirming that $ E_3 $ remains stable for all~$ \tau_1, \tau_2 >0$.  



\subsubsection{A geometric transversality condition.}

Once we have established the existence of imaginary roots for the pseudo-polynomial~\eqref{quasy}, we now provide a geometric characterisation of their crossing of the imaginary axis as a function of the delays $\tau_1,\tau_2>0$.  Given the oscillatory behaviour of the steady-state $E_1$ at $\tau_1 = \tau_2 = 0$, our analysis is focused on this point.  Crucially, $q(0,\boldsymbol{\tau},\mathbf{p}) \neq 0$, as shown by~\eqref{quasy}. Furthermore, the polynomials $p_i(\lambda,\boldsymbol{\tau},\mathbf{p})$, for $i=1,2,3$, defined in  \eqref{p3}-\eqref{p1}, are numerically verified to have no common zeros, and $\deg(p_3(\lambda,\boldsymbol{\tau},\mathbf{p})) > \max\{\deg(p_2(\lambda,\boldsymbol{\tau},\mathbf{p})),\deg(p_1(\lambda,\boldsymbol{\tau},\mathbf{p}))\}$. These conditions ensure that for~$E_1$, there exist values of $\tau_1$ and $\tau_2$ for which the real parts of the roots of $q(\lambda,\boldsymbol{\tau},\mathbf{p})$ lie in the left-hand side of the complex plane; see, for instance, \cite{gu2005stability}.
 
\begin{figure}[t!]
    \centering
    \includegraphics[scale=0.375]{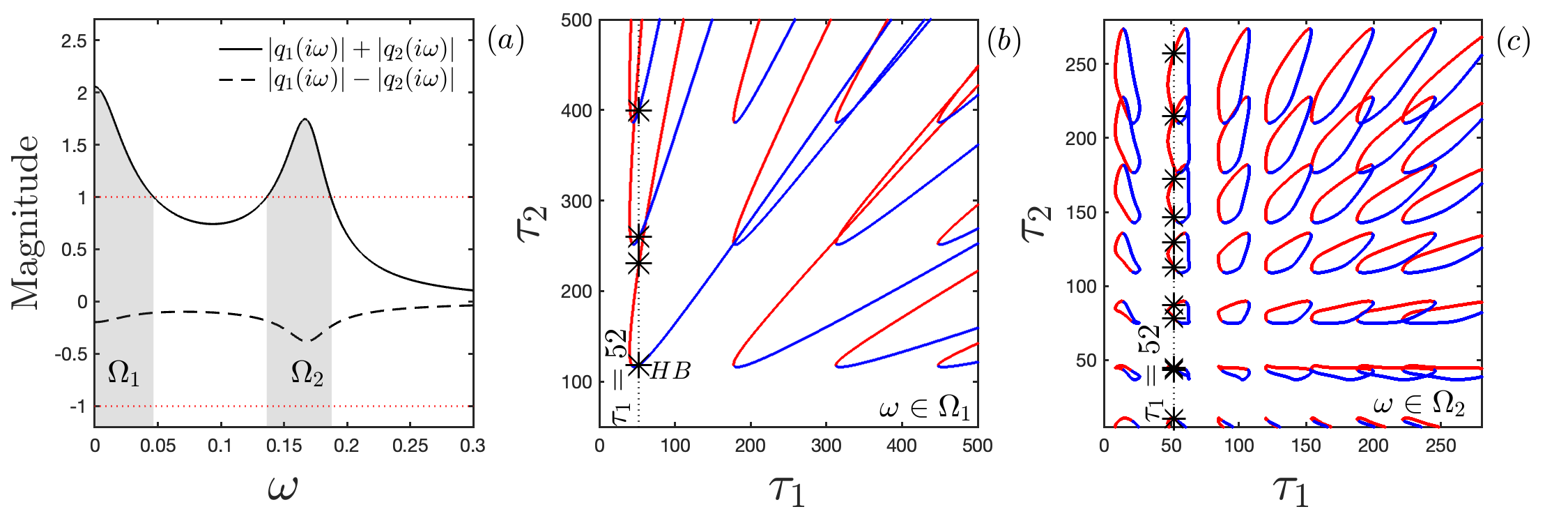}
    \caption{The pseudo-characteristic polynomial \eqref{quasy} possesses at least one purely imaginary root when $\omega$ simultaneously satisfies the geometric conditions \eqref{triang}. This occurs within two distinct frequency intervals: a lower interval, $\Omega_1=[0,0.0466554]$, and a higher interval, $\Omega_2=[0.136335,0.187117]$, indicated by the grey bands in panel~(a). Panels~(b) and~(c) illustrate all positive time delays for which the pseudo-polynomial roots lie on the imaginary axis, parameterised by $\omega$. Specifically, panel~(b) shows trajectories for lower frequencies ($\omega\in\Omega_1$), while panel~(c) displays those for higher frequencies ($\omega\in\Omega_2$). Asterisks mark the $(\tau_1,\tau_2)$ pairs where $\tau_1=52$. The direction in which roots cross the imaginary axis, as a function of time delay variation, is characterised by the transversality parameter~\eqref{transv}: blue curves indicate a crossing from left to right on the imaginary axis, while red curves denote a crossing from right to left.
    }
\label{fig:paramUnified}
\end{figure}

To characterise the intersections, we interpret the three terms in~\eqref{quasySum1} as vectors in the complex plane. The first term has a magnitude of unity and is oriented along the real axis. The remaining two terms are given by $|q_j(i\omega)|\exp(-i\omega\tau_j+i\arg(q_j(i\omega)))$, for $j=1,2$.  Since equation (\ref{quasySum1}) holds, these vector orientations must form a triangle. This geometric constraint is satisfied only for non-zero values of $\omega$ that also fulfill the conditions stated in Proposition 3.1 of \cite{gu2005stability}:
\begin{gather}\label{triang}
1\leq \left|q_1(i\omega)\right|+\left|q_2(i\omega)\right|\,, \quad -1\leq \left|q_1(i\omega)\right|-\left|q_2(i\omega)\right|\leq 1\,.
\end{gather}
These conditions are illustrated in Figure~\ref{fig:paramUnified}(a) at the steady-state $E_1$.  As is depicted there, inequalities in~\eqref{triang} are simultaneously satisfied only within the subintervals $\Omega_1 = [0, 0.0466554]$ and $\Omega_2 = [0.136335, 0.187117]$ for the parameter values given in Table~\ref{table:Ap1}.

Now, as is established in \cite{gu2005stability}, equation (\ref{quasySum1}) yields expressions for the delays, $\tau_1 = \tau_1^{n\pm}(\omega)$ and $\tau_2 = \tau_2^{m\pm}(\omega)$, as functions of $\omega > 0$, given by
\begin{subequations}\label{eq:taus}
\begin{gather}
\tau_1^{n^{\pm}}(\omega) = \frac{\arg(q_1(i\omega))}{\omega} \pm \frac{1}{\omega}\arccos\left(\frac{1+|q_1(i\omega)|^2-|q_2(i\omega)|^2}{2|q_1(i\omega)|}\right) + \frac{(2n-1)\pi}{\omega},\label{taus1}\\[1ex]
\tau_2^{m^{\pm}}(\omega) = \frac{\arg(q_2(i\omega))}{\omega} \mp \frac{1}{\omega}\arccos\left(\frac{1+|q_2(i\omega)|^2-|q_1(i\omega)|^2}{2|q_2(i\omega)|}\right) + \frac{(2m-1)\pi}{\omega},\label{taus2}
\end{gather}
\end{subequations}
where $n$ and $m$ are integers such that $n = n_0^+, n_0^+ + 1, n_0^+ + 2, \dots$ and $m = m_0^+, m_0^+ + 1, m_0^+ + 2, \dots$, respectively.  Here, $n_0^+$, $n_0^-$, $m_0^+$, and $m_0^-$ represent the smallest non negative integers for which $\tau_1^{n_0^{\pm}}$ and $\tau_2^{m_0^{\pm}}$ are non negative. These delay values can be gathered in the set $\mathcal{T}_{\omega,n,m}^{\pm} = \left\{(\tau_1^{n^{\pm}}(\omega),\tau_2^{m^{\pm}}(\omega))\right\} \subset \mathbb{R}_{+}^2$, parameterised by~$\omega>0$, and define $\mathcal{T}_{n,m}^{\pm k} = \bigcup_{\omega \in \Omega_k} \mathcal{T}_{\omega,n,m}^{\pm}$ for $k=1,2$.  Hence, the set $\mathcal{T}=\mathcal{T}^1\cup\mathcal{T}^2$, where 
\begin{gather*}
\mathcal{T}^k=\bigcup_{n,m\in\mathbb{Z}}\left(\mathcal{T}_{n,m}^{+ k}\cup\mathcal{T}_{n,m}^{- k}\right)\cap \mathbb{R}_{+}^2\,, \quad k=1,2\,,
\end{gather*}
consists of all points $(\tau_1,\tau_2) \in \mathbb{R}_+^2$ for which the pseudo-polynomial $q(i\omega,\boldsymbol{\tau},\mathbf{p})$ possesses at least one zero on the imaginary axis; see~\cite{gu2005stability}.  In Figure \ref{fig:paramUnified}, the points $(\tau_1,\tau_2)$ in the $\mathbb{R}_+^2$ plane corresponding to the sets $\mathcal{T}^1$, for $\omega \in \Omega_1$,  and $\mathcal{T}^2$, for $\omega \in \Omega_2 $, can be seen in panels (b) and (c), respectively, for the parameter set values in Table~\ref{table:Ap1}.

To analyse the direction of imaginary axis crossings for the roots of $q(\lambda, \boldsymbol{\tau}, \mathbf{p})$ at points $(\tau_1, \tau_2) \in \mathcal{T}^k$, where $k=1,2$, we consider eigenvalues of the form $\lambda = \sigma + i\omega$.  Specifically, we examine how the real part, $\sigma$, changes as the delays vary.

First, we define the tangent vector to $\mathcal{T}^k$ along the curve of increasing~$\omega$ as $\mathbf{w}_{\theta} = (\partial \tau_1/\partial \omega, \partial \tau_2/\partial \omega)$.  A normal vector pointing towards the left-hand region of the positively oriented curve $\mathcal{T}^k$ is then provided by $\mathbf{v}_{\theta} = (-\partial \tau_2/\partial \omega, \partial \tau_1/\partial \omega)$.

Second, we consider the scenario where a pair of complex conjugate roots of $q(\lambda, \boldsymbol{\tau}, \mathbf{p})$ crosses the imaginary axis into the right-hand side of the complex plane.  In this case, the delay vector $\boldsymbol{\tau} = (\tau_1, \tau_2)$ shifts parallel to the vector $\mathbf{w}_{r} = (\partial \tau_1/\partial \sigma, \partial \tau_2/\partial \sigma)$.  Consequently, if $\boldsymbol{\tau} \in \mathbb{R}_+^2$ lies to the left of a positively oriented orbit $\mathcal{T}^k$, two additional roots of $q(\lambda, \boldsymbol{\tau}, \mathbf{p})$ become unstable when the inner product $\langle \mathbf{w}_{r}\,,\mathbf{v}_{\theta} \rangle > 0$. Thus, as a consequence of the Implicit Function Theorem and Proposition 6.1 of \cite{gu2005stability}, this condition is equivalent to~$\mathcal{C}>0$, where
\begin{gather}\label{transv}
\mathcal{C}:=\text{Im}\left(q_1(i\omega)q_2(-i\omega)e^{i\omega(\tau_2-\tau_1)}\right)\,,
\end{gather}
which is known as {\em geometric transversality parameter} as it provides a criterion for determining the crossing direction of the roots. 

The parameter $\mathcal{C}$, defined in~\eqref{transv}, characterises the direction of root crossing and, consequently, the type of Hopf bifurcation.  Figure~\ref{fig:paramUnified}(b)-(c) visualises the delays $(\tau_1, \tau_2)$ categorised by this parameter.  Red curves delineate regions where $\mathcal{C}<0$, while blue regions represent $\mathcal{C} > 0$.  Each intersection of a root of~$q(\lambda,\boldsymbol{\tau},\mathbf{p})$ with the imaginary axis corresponds to a Hopf bifurcation.  Specifically, a {\em supercritical} Hopf bifurcation occurs when the root crosses from left to right (i.e., $\mathcal{C}>0$), while a {\em subcritical} Hopf bifurcation occurs for a crossing from right to left (i.e., $\mathcal{C}<0$).  {
This leads directly to the following 
\begin{prop}\label{prop:transv}
Under conditions~\eqref{triang}, the geometric transversality parameter~\eqref{transv} dictates the criticality of the Hopf bifurcation:
\begin{enumerate*}[label=(\alph*)]
	\item it is supercritical if $\mathcal{C}>0$,
	\item and subcritical if $\mathcal{C}<0$.
\end{enumerate*}
\end{prop}
}The likelihood of these bifurcations increases with the difference between the delays, $|\tau_1 - \tau_2|$.  Moreover, the direction of root crossing exhibits a quasi-periodic behaviour, as illustrated in Figure~\ref{fig:paramUnified}(b)-(c).  These results further suggest a dependence of the bifurcation type on the frequency.  Specifically, the emergence of $\alpha$- and $\omega$-limit cycles is observed across a range of oscillatory frequencies. These limit cycles manifest at both lower frequencies (Figure~\ref{fig:paramUnified}(b)) and elevated frequencies (Figure~\ref{fig:paramUnified}(c)), a phenomenon associated with an alternating sequence of criticality transitions. This observation underscores the nuanced relationship between oscillatory dynamics and system criticality within the explored parameter set values.


\section{Numerical bifurcation analysis of time delay QS system.}
\label{sec:third}%

{As established in the preceding section, the introduction of two delays significantly alters the system's dynamics, inducing self-sustained oscillations through Hopf bifurcations. Such bifurcations associated with time delay combinations, specifically with fixed $\tau_1$ and varying $\tau_2$, are visually represented by the collection of asterisks along the dotted lines ($\tau_1 = 52$) in Figures~\ref{fig:paramUnified}(b) and~\ref{fig:paramUnified}(c). These figures demonstrate that both subcritical (red) and supercritical (blue) Hopf bifurcations occur in both lower ($\omega\in\Omega_1$) and higher ($\omega\in\Omega_2$) frequency ranges. To further investigate stability changes due to the introduction of time delays, we fixed the parameter $\tau_1$ as previously mentioned and slowly varied $\tau_2$. A bifurcation diagram is computed by using DDE-BIFTOOL~\cite{engelborghs2002numerical} by extending a stability branch in the neighbourhood of equilibrium $E_1$ and numerically locating Hopf bifurcations within the range $\tau_2\in[2,200]$. }

\begin{figure}[t!]
\centering
\includegraphics[scale=0.323]{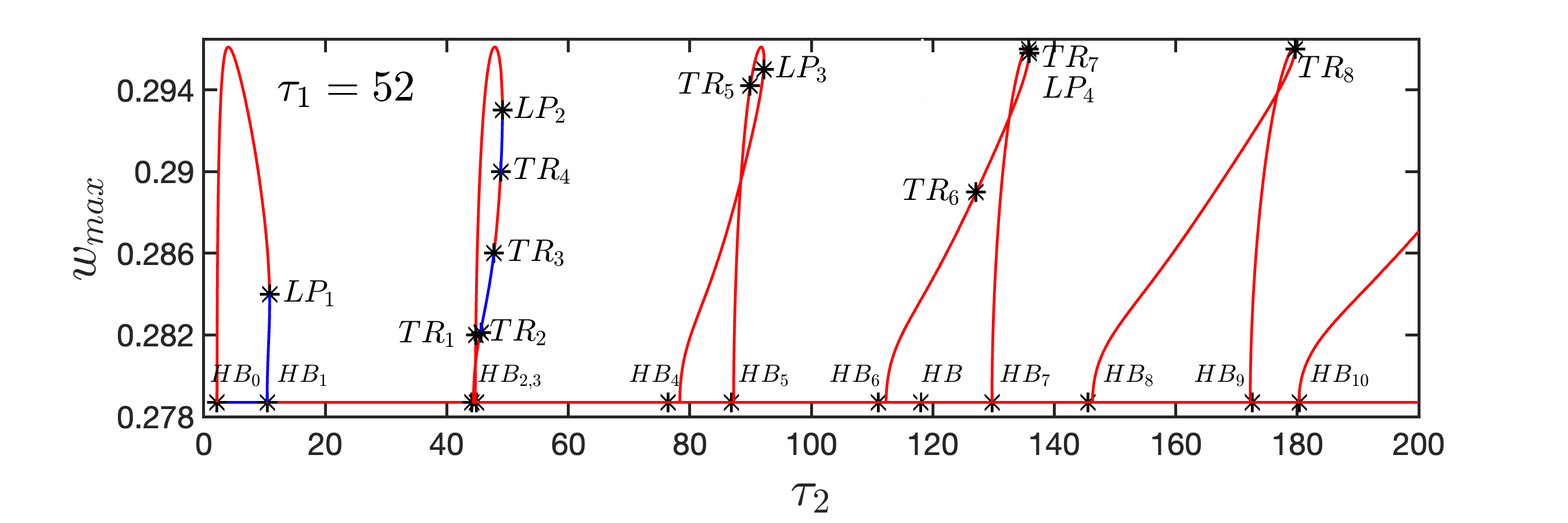}
\caption{Extended stability and periodic orbit continuation. The system's stability and periodic orbit extensions with respect to the varying delay parameter $\tau_2$, with $\tau_1=52$ held fixed. The primary Hopf bifurcation $HB_0$ and ten secondary Hopf bifurcations ($HB_i$, for $i = 1, \dots, 10$) are identified, and their locations and criticality satisfactorily correspond with those predicted in Figure~\ref{fig:paramUnified}(c). The point labelled $HB$, situated between $HB_6$ and $HB_7$, corresponds to the first $HB$ point (not extended) shown in Figure~\ref{fig:paramUnified}(b). Periodic orbit branches emanate from each Hopf point. Stable branches, representing either periodic or steady-state orbits, are indicated in blue, while unstable branches are shown in red. Asterisks mark detected torus $TR$ and fold $LP$ bifurcation points along these periodic orbit branches.}
\label{figBif}
\end{figure}

{
The bifurcation diagram in Figure~\ref{figBif} validates the computations presented in Figure~\ref{fig:paramUnified}(b) and~2(c) regarding the location of Hopf points within the range $\tau_2\in[0,200]$. We observed slight numerical discrepancies between the identified Hopf bifurcation points in Figures~\ref{fig:paramUnified} and~\ref{figBif}, though these remained within acceptable numerical errors. Our analysis of criticality will now focus on the Hopf bifurcations shown in Figure~\ref{fig:paramUnified}(c), as the diversity of dynamics becomes increasingly complex in the higher frequency range $\omega\in\Omega_2$.

Specifically, the criticality of the bifurcations in Figure~\ref{fig:paramUnified}(c), predicted with the aid of the geometric transversality parameter~\eqref{transv}, is consistent with the stability results presented in Figure~\ref{figBif}. That is, $HB_0$ corresponds to a subcritical Hopf bifurcation, $HB_1$ to a supercritical Hopf bifurcation, and the remaining $HB_i$ for $i=2,...,10$ all denote subcritical Hopf bifurcations.
}


Following the identification of the $\tau_2$ range associated with Hopf bifurcations, numerical continuation of equilibrium and periodic branches was performed using DDE-BIFTOOL. Figure~\ref{figBif} displays these branches, including their stability properties (stable branches in blue, unstable branches in red), in a neighbourhood of the equilibrium point $E_1$, parameterised by $\tau_2 \in [2, 200]$ with $\tau_1 = 52$.  The Hopf bifurcations marked with asterisks in Figure~\ref{fig:paramUnified}(c) were successfully located using DDE-BIFTOOL and correspond to the $HB_i$ points in Figure~\ref{figBif}. Note that $HB_0$ is not shown in Figure~\ref{fig:paramUnified}(c) due to the figure's scale.  The $HB$ point in Figure~\ref{figBif} (between $HB_6$ and $HB_7$) corresponds to the first $HB\approx117.3664$ in Figure~\ref{fig:paramUnified}(b),{ which we did not perform a numerical continuation for this particular case, as our focus is directed towards the higher frequency regime characterised by $\omega\in\Omega_2$}.  Slight numerical discrepancies were observed between the Hopf bifurcation points identified in Figures~\ref{fig:paramUnified} and~\ref{figBif}.

Furthermore, as illustrated in Figure~\ref{figBif}, the analysis of periodic solution bifurcations revealed several torus ($TR$) and fold ($LP$) bifurcations.  However, neither period-doubling bifurcations, characteristic of the Ruelle--Takens--Newhouse route to chaos~\cite{newhouse1978occurrence}, nor homoclinic bifurcations, indicative of the Shilnikov homoclinic chaos mechanism (see, e.g.,~\cite{seb05,gonchenko97}), were detected within the explored parameter regime.  This absence suggests that, if a chaotic regime exists, it likely arises through a different bifurcation scenario.

\begin{figure}[t!]
\centering
\includegraphics[scale=0.45]{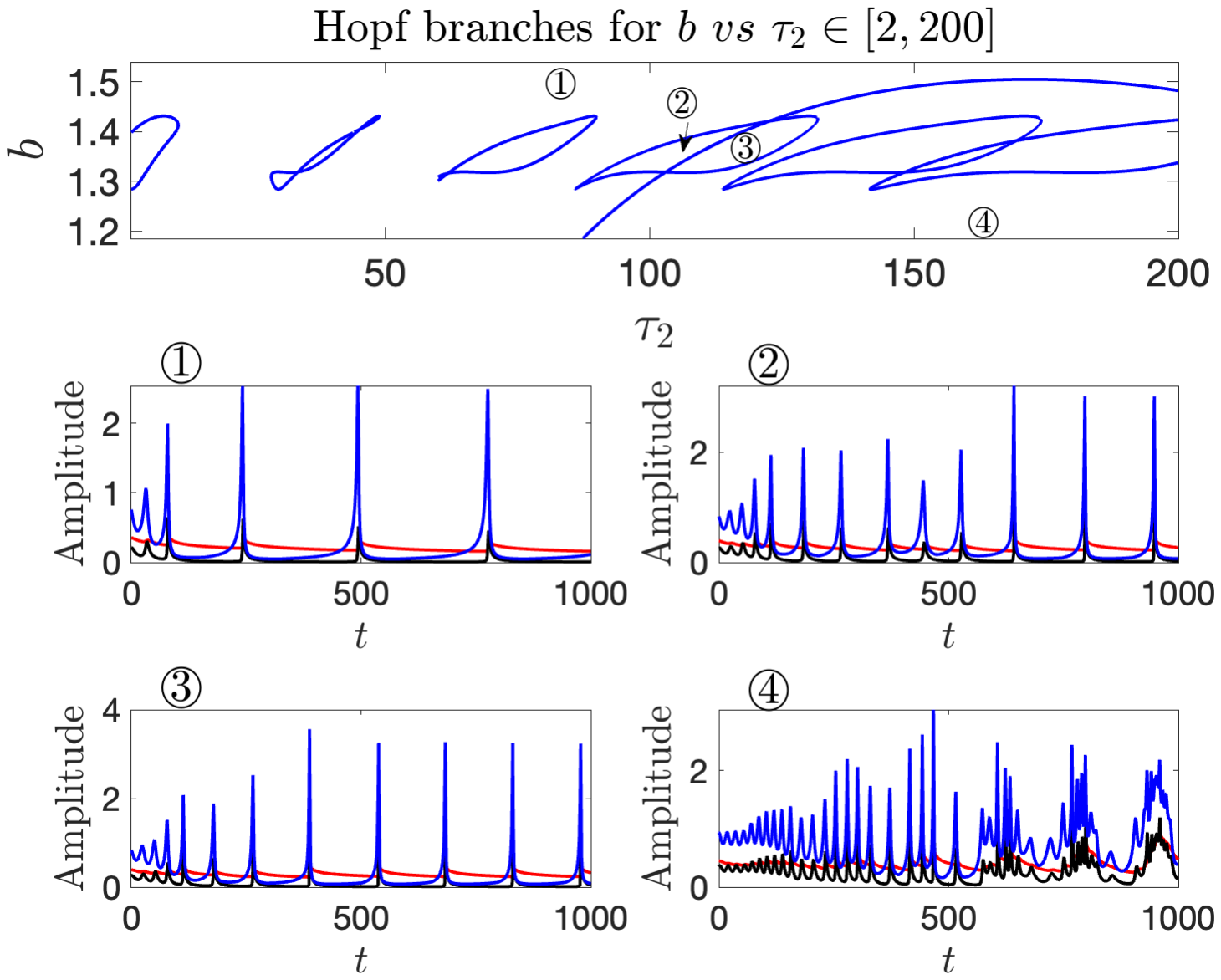}
\caption{Top Panel: this panel presents a two-parameter continuation of Hopf bifurcations within the range $\tau_2 \in [0,200]$, with $\tau_1=52$ fixed. The emergence of intersecting islands as $\tau_2$ increases suggests the onset of complex oscillations. Panels (1)-(4): to further investigate the complexity in the time evolution scenario, the temporal dynamics of the system variables $u(t)$, $v(t)$, and $w(t)$ (red, black, and blue lines, respectively) are computed for selected parameter values: (1) $\tau_2 = 80$, $b = 1.5$; (2) $\tau_2 = 105$, $b = 1.36$; (3) $\tau_2 = 115$, $b = 1.36$; and (4) $\tau_2 = 160$, $b = 1.2$. Other parameter values are as in Table~\ref{table:Ap1}. While panels (1) through (3) display qualitatively similar temporal dynamics, panel (4) indicates the onset of complex, non-periodic oscillations.
}
\label{Fig:inout}
\end{figure}

{To further investigate the presence of complex oscillations on the system, a} two-parameter continuation was then performed, varying $b$ and $\tau_2$ for each Hopf bifurcation point within the range $\tau_2 \in [2, 200]$.  The parameter~$b$, related to the decay rate of motile to static bacteria, was chosen as the secondary bifurcation parameter due to its critical role in influencing the interplay of self-sustained oscillatory dynamics, as discussed in~\cite{harrisvf}. The results of this analysis are presented in the top panel of Figure~\ref{Fig:inout}.  The two-parameter continuation of the Hopf branches reveals the formation of intersecting islands as $\tau_2$ increases, suggesting the potential for complex oscillatory behaviour. To shed light on this possibility, specific values of~$b$ and~$\tau_2$ were selected to explore key system behaviours over time. These values are: $\tau_2 = 80$, $b = 1.5$ (point 1); $\tau_2 = 105$, $b = 1.36$ (point 2); $\tau_2 = 115$, $b = 1.36$ (point 3); and $\tau_2 = 160$, $b = 1.2$ (point 4).  The corresponding time series are shown in panels 1-4 of Figure~\ref{Fig:inout}. While the temporal dynamics depicted in panels 1 through 3 are qualitatively similar, panel 4 suggests the emergence of complex, non-periodic oscillations. This behaviour may result from the combined effect of small $b$ values and large $\tau_2$ values, given that $\tau_1$ remains fixed at 52.

\section{Intermittent chaos in time delay QS system}
\label{sec:four}
\begin{figure}[t!]
    \centering
    \includegraphics[scale=0.46]{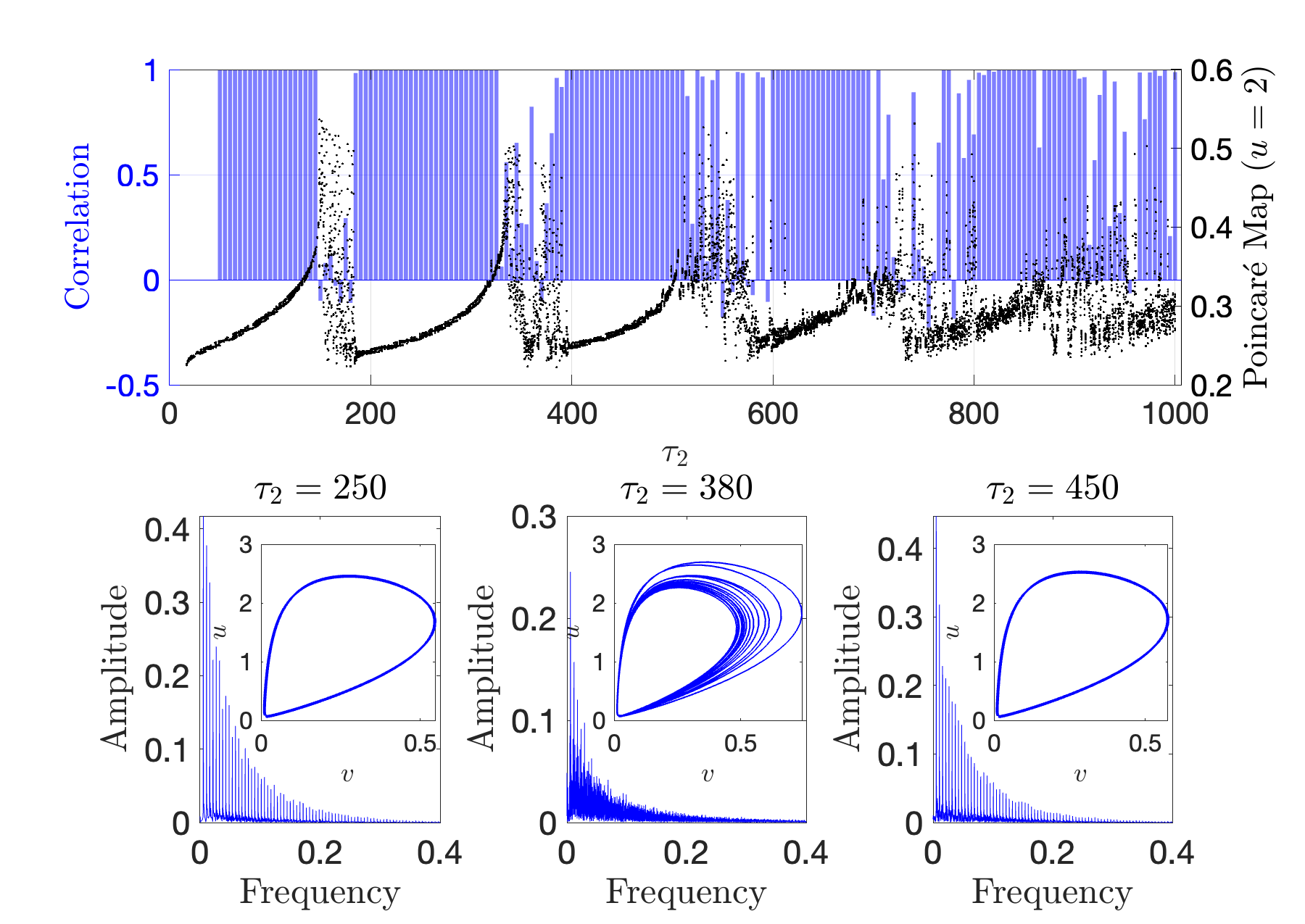}
    \caption{Upper row: Poincaré map computed on the cross-section $u=2$ (black dots) after transient time, as $\tau_2$ is slowly varied; Pearson correlation coefficient (blue bars) measuring the correlation between two initially nearby $w$-component orbits. Lower row: frequency-amplitude plots and projections of representative orbits onto the ($u$, $v$) plane for three key distinct values of $\tau_2$; the central panel depicts a quasi-periodic orbit, while the left-hand and right-hand panels show stable periodic orbits. Other parameter values as in Table~\ref{table:Ap1}.}
    \label{fig:chaos1.4}
\end{figure}

In the preceding section, the existence of periodic self-sustained oscillations was established for specific parameter values, as illustrated in Figure~\ref{Fig:inout}, panels~1-3. These periodic oscillations correspond to stable fixed points of the Poincaré map.  Furthermore, as the time delay $\tau_2$ increases, the stability of these oscillations is altered, leading to a deformation of the oscillatory pattern, as observed in panel 4. This deformation suggests the emergence of complex, potentially non-periodic, dynamics.

To gain a comprehensive understanding of the oscillatory features governed by critical bifurcations, the Poincaré map of the time-delayed QS system~\eqref{eq:qsd} was computed over a substantial range of $\tau_2$ values, while holding $\tau_1$ constant. The first row of Figure~\ref{fig:chaos1.4} depicts the intersections of the state variable $u$ with the cross-section $u=2$ (black dots)---effectively, the Poincaré map---after transient time, as $\tau_2$ is slowly increased within the interval $\tau_2 \in [2, 1000]$ for fixed values of $\tau_1 = 52$ and $b = 1.4$.  As can be seen there, intervals of stable periodic solutions and complex quasi-periodic oscillations are interspersed until $\tau_2$ reaches a value sufficient to completely disrupt the periodic dynamics; see Figure~\ref{fig:chaos1.4}, upper panel, $\tau_2\approx700$. Beyond this point, the non-periodic oscillations suggest a chaotic regime, which is further explored in the bottom row of Figure~\ref{fig:chaos1.4} through frequency-amplitude plots and projections of distinguished orbits onto the $(u,v)$ phase plane.  That is, for~$\tau_2 = 250$, the Poincaré map indicates a stable periodic behaviour { as can be seen from the trend forming the black curve}, which is corroborated by the well-defined orbit in the $(u,v)$ phase plane and the discrete frequency spectrum shown in the left-hand panel of the bottom row.  When~$\tau_2 = 380$, the dispersed Poincaré points and the dense orbit in the phase plane, coupled with the continuous frequency spectrum in the central panel, strongly indicate a chaotic regime. Finally, at $\tau_2 = 450$, the system returns to a stable periodic window, and the conclusions drawn for the case of $\tau_2=250$ apply.

\begin{figure}[t!]
    \centering
    \includegraphics[scale=0.21]{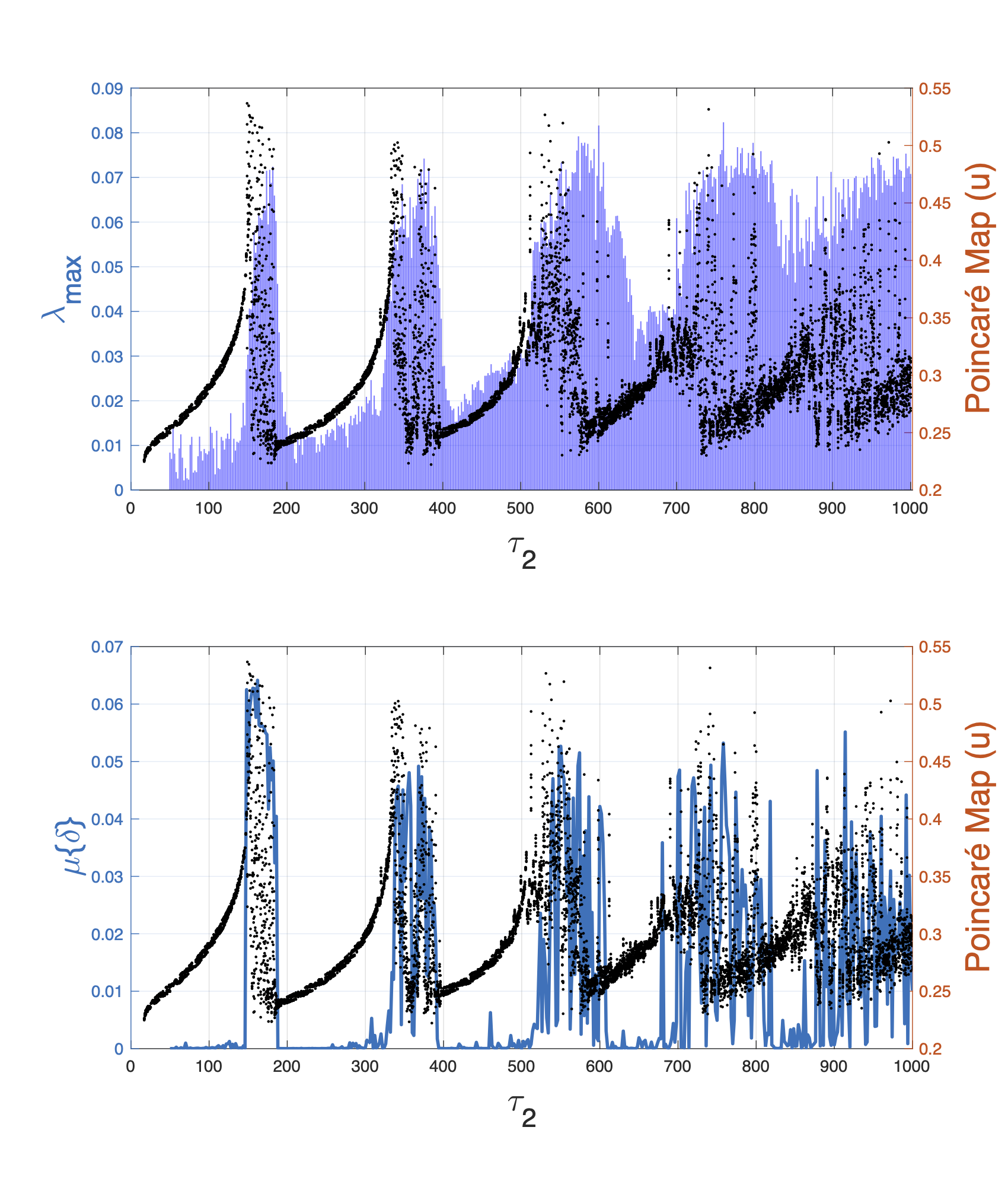}
    \caption{Initial conditions sensitivity for varying delay parameter~$\tau_2$. Upper panel: the largest Lyapunov exponent $\lambda_{\textrm{max}}$ is shown as a function of $\tau_2$. Positive values (blue shaded regions) indicate chaotic regimes, while values near zero denote self-sustained periodic oscillations. Lower panel: the mean curve separation~$\mu\{\delta\}$ (initial perturbation $\delta_0 \approx 10^{-7}$), in blue, illustrates divergence in the chaotic windows and convergence to naught in non-chaotic intervals after $t = 17.5 \times 10^4$. The accompanying Poincar\'e map of $u$ (black dots) in both panels visually confirms the system's transitions between chaotic sets and stable limit cycles, displaying the intermittent behaviour.}
    \label{Fig:lyapunov_analysis}
\end{figure}

To further validate the observed chaotic behaviour, the Pearson correlation coefficient~\cite{walpole2012probabilidad} was computed and is displayed as blue bars in Figure~\ref{fig:chaos1.4}, superimposed on the Poincaré map.  Given two column vectors $x \in \mathbb{R}^{n \times 1}$ and $y \in \mathbb{R}^{n \times 1}$, with their respective averages $\overline{x} = \sum_{i=1}^n x_i / n$ and $\overline{y} = \sum_{i=1}^n y_i / n$, the Pearson linear correlation coefficient is given by
\begin{gather}\label{eq:pearson}
	r_{x,y} = \dfrac{\sum\limits_{i=1}^n (x_i - \overline{x})(y_i - \overline{y})}{\sqrt{\sum\limits_{i=1}^n (x_i - \overline{x})^2} \sqrt{\sum\limits_{i=1}^n (y_i - \overline{y})^2}}.
\end{gather}
This coefficient ranges from -1 to 1, where 1 indicates perfect positive correlation, -1 perfect negative correlation, and 0 no correlation. In this context, the correlation was calculated between two discretised orbits of the $w$ state-component, initialised with a difference of approximately $10^{-9}$ units. As shown in Figure~\ref{fig:chaos1.4}, the blue bars reveal strong correlation between $w$-orbits within the periodic regimes, while the correlation collapses to naught in the windows where chaotic behaviour dominates. This characteristic pattern is indicative of {\em intermittent chaos}, a phenomenon first described by Pomeau and Manneville in their analysis of the Lorenz system~\cite{manneville1979intermittency}.  Intermittent chaos is a well-established phenomenon in time-delayed regulatory gene circuits, often attributed to the inherent delays in intracellular processes~\cite{suzuki2016periodic}.

{
To further analyse this phenomenon, we compute the largest Lyapunov exponent $\lambda_{\textrm{max}}$ using the method proposed by Rosenstein {\em et al.}~\cite{rosenstein1993practical}, as implemented in the MATLAB routine \texttt{lyapunovExponent}~\cite{matlab}. The top panel of Figure~\ref{Fig:lyapunov_analysis} depicts $\lambda_{\textrm{max}}$ as a function of $\tau_2$. It is worth noting that $\lambda_{\textrm{max}} > 0$ even in some seemingly non-chaotic regions. This observation may be a consequence of the phenomenon reported by Grantham {\em et al.} in~\cite{grantham1993} and Stefanski {\em et al.} in~\cite{stefanski2010}, where positive finite-time Lyapunov exponents (FTLEs) characterise the initial, transient behaviour of a system before it settles onto a stable limit cycle. Their work demonstrated an inverse relationship: larger positive FTLEs during the chaotic transient correspond to shorter ``rambling times'', meaning the system escapes the chaotic phase more quickly to reach its periodic orbit. Conversely, smaller positive FTLEs indicate a more prolonged period of transient chaos before convergence to the limit cycle. This implies that even when a system ultimately approaches a stable non-chaotic attractor, FTLEs can quantify the intensity and persistence of its preceding chaotic transient. Moreover, it is important to recognise that numerically computed positive FTLEs can emerge as numerical artefacts of discretisation in systems theoretically exhibiting stable limit cycles. As demonstrated by De Markus {\em et al.} in~\cite{demarkus2000}, the integration step size can induce ``computational chaos'' where FTLEs become positive for periodic systems, or conversely, lead to numerically induced periodicity in truly chaotic ones. This highlights the crucial role of numerical methods in shaping observed dynamics, suggesting that positive FTLEs in stable systems often indicate numerical errors or transient chaos rather than sustained chaotic behaviour, for instance.

\begin{figure}[t!]
    \centering
    \includegraphics[scale=0.151]{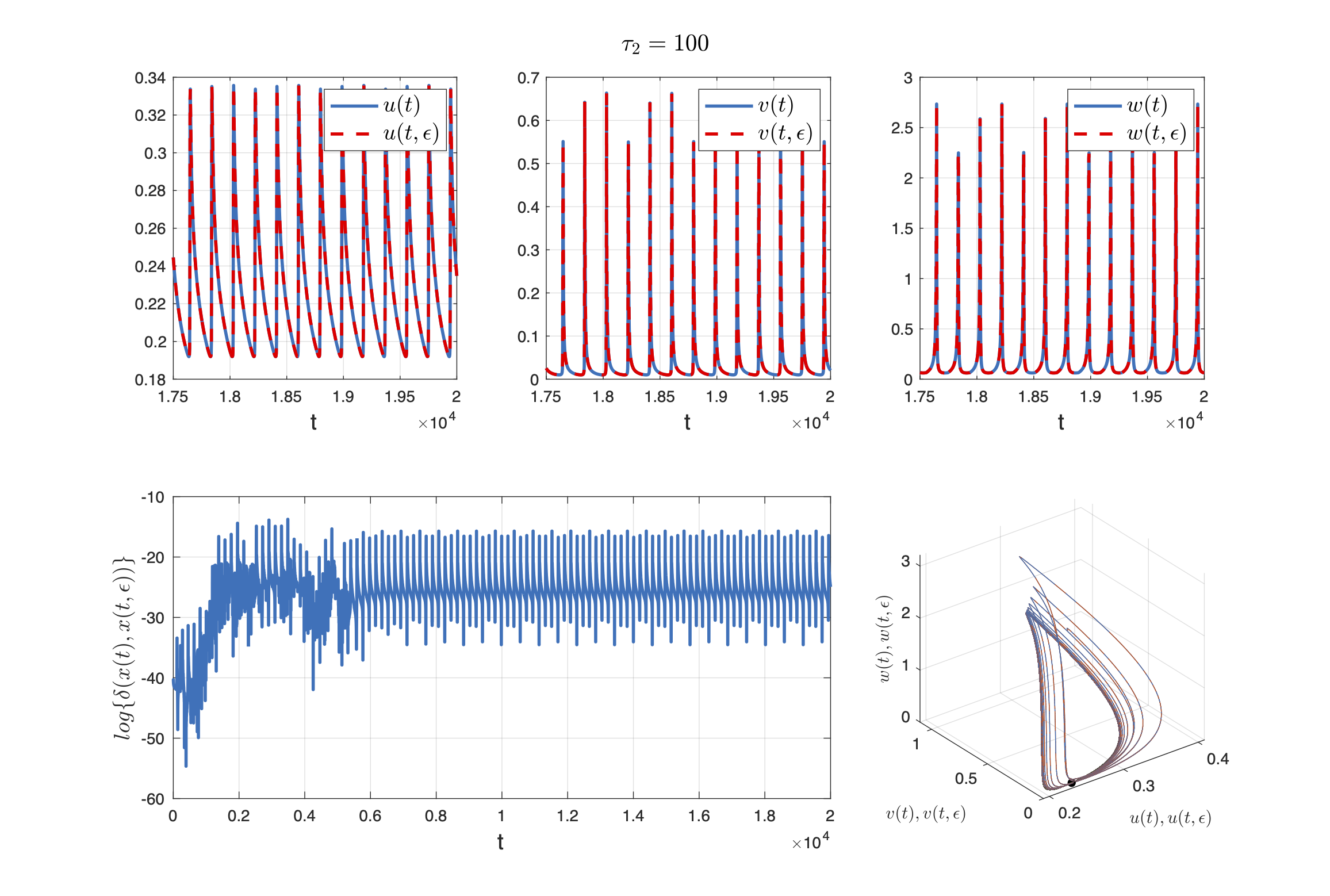}\\[-1ex]
    \includegraphics[scale=0.151]{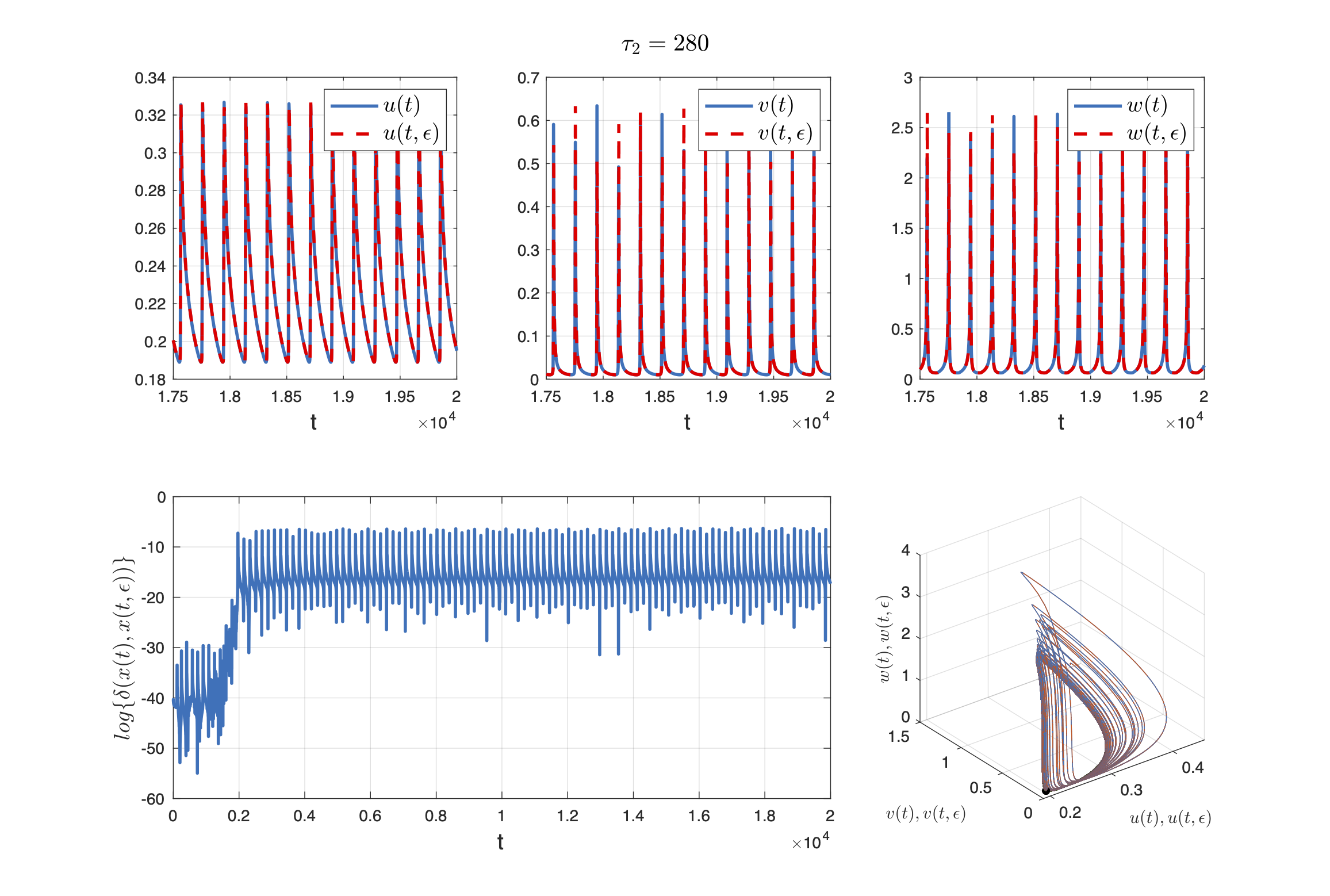}
    \caption{
        Non-chaotic regime for $\tau_2 = 100,280$. Upper panels: time series of state variables exhibit stable self-sustained oscillations for $x(t)$ and $x(t, \epsilon)$. Bottom left-hand panel: log-distance between perturbed trajectories. Bottom right-hand panel: the two phase-space orbits further confirms a stable limit cycle. Other parameter values as in Table~\ref{table:Ap1}.
    }
    \label{Fig:SimTau2_100_280}
\end{figure}




To properly characterise our findings, $\lambda_{\textrm{max}}$ is presented as $\tau_2$-dependent alongside the mean trajectory separation distance $\mu\{\delta\}$ in the bottom panel of Figure~\ref{Fig:lyapunov_analysis}. This quantity represents the average distance between two orbits after a sufficiently long time, specifically $t = 17.5 \times 10^4$, with an initial perturbation magnitude of $\delta_0 \approx 10^{-7}$ between their initial conditions. To enable a comprehensive comparison of stable limit cycles and chaotic windows across both panels, a Poincar\'e map is superposed.

Observe that $\lambda_{\textrm{max}}$ clearly increases by nearly an order of magnitude as the system becomes chaotic. Conversely, it decreases when stable limit cycles are dominant, as illustrated in the top panel of Figure~\ref{Fig:lyapunov_analysis}. This behaviour is further corroborated by~$\mu\{\delta\}$, which attains significantly larger values in chaotic regions, consistent with exponential divergence, while dropping significantly during periodic oscillations, as shown in the bottom panel of Figure~\ref{Fig:lyapunov_analysis}. The interplay between $\lambda_{\textrm{max}}$ and $\mu\{\delta\}$, combined with the corresponding Poincar\'e map, which exhibits no presence of accumulation points in chaotic regions, and the Pearson linear correlation coefficient given by equation~\eqref{eq:pearson}, confirms the presence of intermittent chaos.

\begin{figure}[t!]
    \centering
    \includegraphics[scale=0.151]{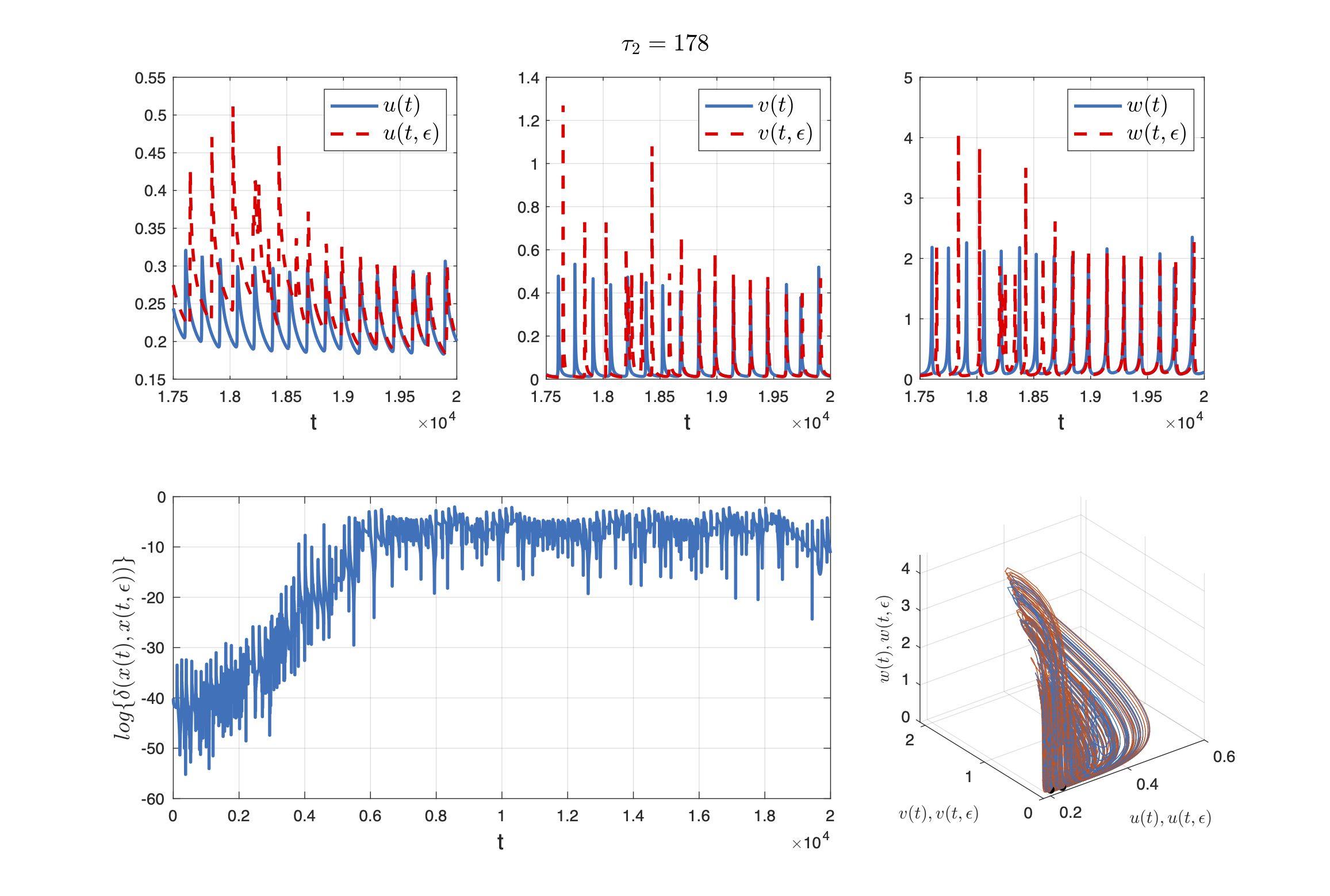}\\[-1ex]
    \includegraphics[scale=0.151]{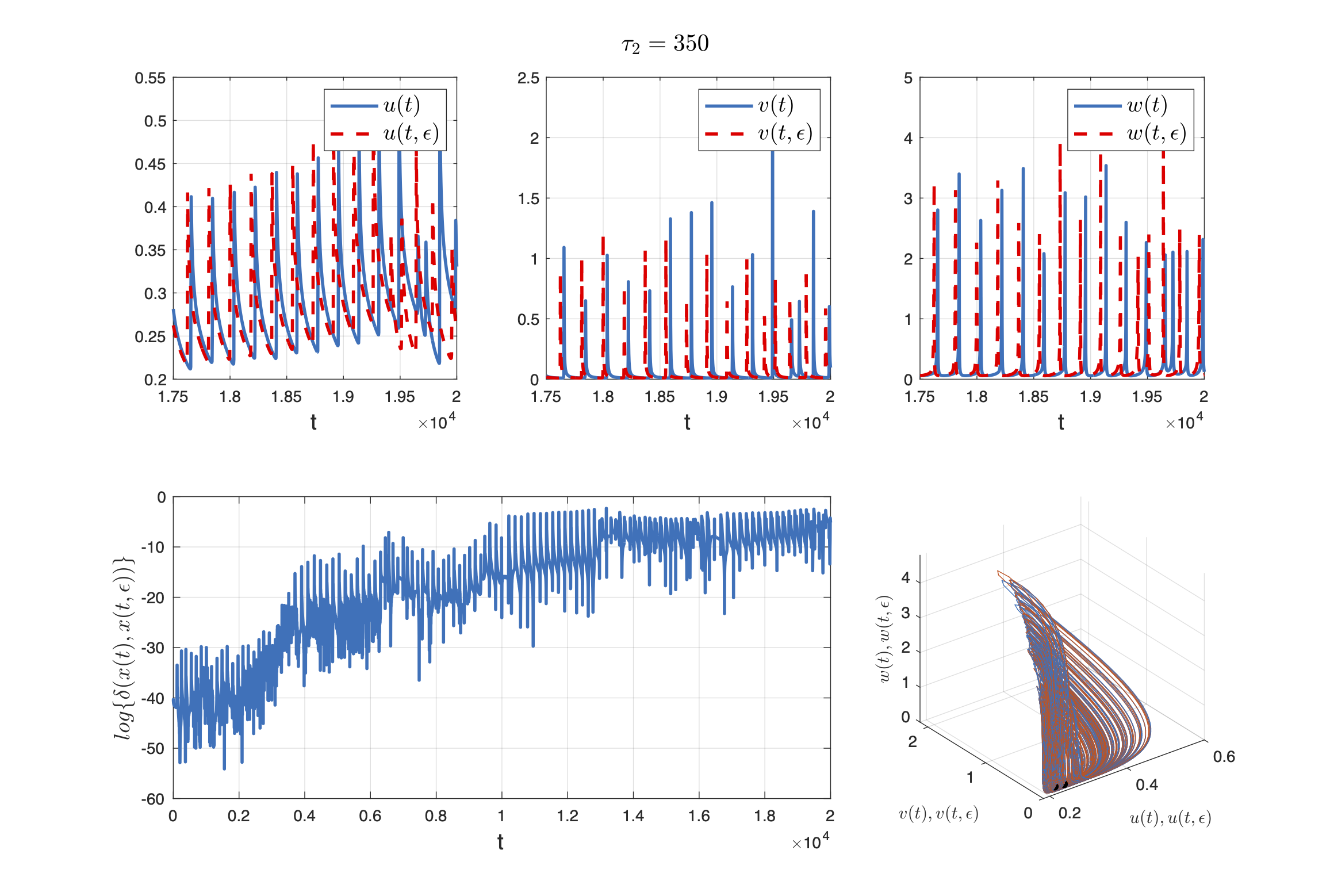}\\[-1ex]
    \caption{
        Chaotic regime for $\tau_2 = 178,350$. Top row: time series of the state variables exhibit quasi-periodic behaviour for $x(t)$ and $x(t, \epsilon)$. Bottom left- and right-hand panels: log-distance between perturbed trajectories, and the two phase-space orbits further show initial conditions sensitivity, respectively. Other parameter values as in Table~\ref{table:Ap1}.
    }
    \label{Fig:SimTau2_178_350}
\end{figure}


To further illustrate these observations, Figures~\ref{Fig:SimTau2_100_280} and \ref{Fig:SimTau2_178_350} depict the system's orbits for selected values of $\tau_2$. Here, $\delta(x(t), x(t, \epsilon))$ denotes the Euclidean distance between a reference orbit $x(t):=(w(t),v(t),u(t))$ (initial state $x(0)$) and a perturbed orbit $x(t, \epsilon):=\left(w(t, \epsilon),v(t, \epsilon),u(t, \epsilon)\right)$ (initial condition $x(0) + \epsilon$, where $0<\|\epsilon\| \ll 1$). For $\tau_2 =100, 280$, where the Poincar\'e map indicates stable self-sustained oscillations, $\delta(x(t), x(t, \epsilon))$ remains bounded, confirming non-chaotic dynamics, as shown in Figure~\ref{Fig:SimTau2_100_280}. In contrast, for $\tau_2=178, 350$, $\delta(x(t), x(t, \epsilon))$ exhibits exponential-like growth during the transient phase, consistent with chaotic behaviour. Notice that, in non-chaotic regimes, $\delta(x(t), x(t, \epsilon))$ stabilises at significantly low values post-transient compared to chaotic cases where the stabilisation is not clearly observed, further validating the computation of an approximation of the largest Lyapunov exponent, as illustrated in Figure~\ref{Fig:SimTau2_178_350}.
}

The comprehensive numerical simulations presented here provided further compelling evidence of the system's richness dynamical repertoire.  Such simulations not only corroborated the analytical predictions regarding the presence of self-sustained periodic oscillations, but also revealed the presence of intermittent chaotic behaviour as the delay parameters slowly vary.  The intermittent nature of this chaotic regime, characterised by distinct phases of laminar flow punctuated by bursts of chaotic activity, strongly suggests a complex interaction between the destabilised steady-state and the emerging oscillatory modes.  That is, the scheme that found and illustrated {in Figures~\ref{fig:chaos1.4} and~\ref{Fig:lyapunov_analysis}} corresponds to Type-II intermittency, within the Pomeau--Manneville classification design, which represents a distinct pathway to chaotic dynamics characterised by the irregular, intermittent switching between laminar phases and bursts of chaotic activity.  A key distinguishing feature of Type-II intermittency is its association with a subcritical Hopf bifurcation.  Specifically, as a system parameter traverses a critical value, a complex conjugate pair of eigenvalues associated with a stable periodic orbit crosses the imaginary axis, destabilising such orbit. The complex monodromic eigenvalues at the bifurcation point are of the form $\mu=\left(1+{\rho}\right)e^{i\theta}$, where the local Poincare map can be described, in polar coordinates, by its normal form 
\begin{gather*}
	\left\{\begin{array}{l}
		r_{n+1}=\left(1+{\rho}\right)r_n+ar^3_n\,,\\[1ex]
		\theta_{n+1}=\theta_n+c+br_n^2\,,
	\end{array}\right.
\end{gather*} 
where $a,b,c\in\mathbb{R}$ are constant, and Type-II intermittency occurs for ${\rho>0}$, see~\cite{elaskar}. This destabilisation gives rise not to a global chaotic attractor, but rather to a {\em repeller} in the vicinity of the now-unstable periodic orbit.  The dynamics near this repeller are inherently chaotic, due to its unstable nature provided by positive Lyapunov exponents, {here described by means of Proposition~\ref{prop:transv} and confirmed by computations shown in Figures~\ref{fig:paramUnified}, \ref{figBif},~\ref{fig:chaos1.4}~and~\ref{Fig:lyapunov_analysis}}. Orbits are intermittently ``injected'' into the neighbourhood of this repeller, leading to a burst of chaotic behaviour.  Following this excursion into the chaotic region, the orbit is subsequently re-injected into the phase space region previously occupied by the stable periodic orbit, resulting in a return to laminar periodic flow as is illustrated in Figure~\ref{fig:chaos1.4}, upper row, {and both panels of Figure~\ref{Fig:lyapunov_analysis}}.  The average durations of these laminar phases, $\bar{l}$, are statistically distributed, typically exhibiting a power-law scaling with the distance of the bifurcation parameter from its critical value, which is $\bar{l}\sim{\rho}^{-1/2}$, see~\cite{elaskar}.  The essential mechanism underlying this phenomenon is the interplay between the destabilised periodic orbit (and its associated repeller) and the reinjection process that governs the transitions between laminar and chaotic phases. 

\section{Concluding Remarks}
\label{sec:five}

{This study delves into the influence of varied time delays on the dynamic behaviour of a QS-inspired system, employing a nonlinear DDE framework. Specifically, we model a system with motile and static bacterial sub-populations, each potentially exhibiting distinct response times to autoinducer signals. This necessitates the incorporation of two independent delay parameters, $\tau_1>0$ and $\tau_2>0$, which represent the intrinsic temporal lags associated with the intricate biochemical processes underlying QS, such as signal production and cellular response.

Our analysis begins by explicitly introducing these two distinct delays into a well-established activator-inhibitor-like framework, effectively representing the dynamic interplay between motile (activator) and static (inhibitor) bacterial sub-populations. We then conduct a thorough analysis of the existence and local stability properties of the system's steady-states. Upon deriving the pseudo-characteristic polynomial and examining the conditions for purely imaginary eigenvalues, we identify the potential for Hopf bifurcations as the delay parameters are slowly varied. Our findings indicate that the presence of multiple delays, particularly when their magnitudes differ significantly, can substantially alter the system's stability and promote the emergence of complex nonlinear oscillatory behaviour. We observe that steady-states $E_1$ and~$E_2$ can undergo Hopf bifurcations, potentially losing stability as the difference between the delays, $h=\tau_1-\tau_2$, increases. In contrast, steady-state~$E_3$ tends to maintain its stability across the explored range of delay values. The observed disparity in the magnitude of $h$ required to induce instability in~$E_1$ versus $E_2$ suggests a complex and intricate interaction between the two distinct delays and the inherent dynamic properties of each steady-state.

Our results underscore the critical importance of considering the inherent temporal dynamics within biological QS systems for accurate modelling. Upon incorporating distinct time delays for motile and static bacteria---representing the characteristic timescales of signal production and cellular response---we reveal their influence on system behaviour. The observed instabilities, the potential for oscillatory behaviour via Hopf bifurcations, and the varied stability of steady-states $E_1$, $E_2$, and $E_3$ under different delay conditions emphasise that the intrinsic timescales of biological processes are crucial factors governing QS dynamics. The difference in delays, characterised by $h$, emerges as a key parameter influencing these dynamics, highlighting the necessity of accounting for  temporal heterogeneity in models of microbial communication to capture realistic behaviours.

The analysis presented here offers novel insights into the role of time delays in shaping the dynamics of QS systems. Upon explicitly incorporating distinct delays for different state components, we illustrate how temporal factors can significantly influence system stability and potentially give rise to a range of complex dynamic behaviours, including  intermittent chaos. The emergence of intermittent chaos in our QS-inspired model, particularly when compared to existing literature on chaotic dynamical features in similar systems (briefly addressed in Section~\ref{sec:intro}), represents a fairly novel route to chaos in such systems.

Intermittent chaos has been a rare and recent discovery within the realm of QS models. While a limited number of studies have explicitly identified and explored this particular dynamical regime, the work by Zakharova {\em et al.}~\cite{zakharova2022} demonstrated its presence in a model of bacterial QS, linking its emergence to specific parameter ranges governing signal production and reception. Complementing this, Singh {\em et al.}~\cite{singh2015} similarly reported intermittent chaotic dynamics within a different QS model, highlighting its potential implications for understanding the robustness and adaptability of bacterial communication networks.

A key distinction between these two foundational works lies in their focus. In~\cite{singh2015}, the authors focused on the general emergence of intermittent chaotic behaviour from the inherent nonlinearities and feedback mechanisms within a broad QS framework, showcasing the possibility of such complex dynamics without pinpointing a specific driving parameter. In contrast, the results reported in~\cite{zakharova2022} specifically gathers the role of time delays, demonstrating how these temporal lags, characteristic of biochemical processes in QS, may act as a crucial bifurcation parameter directly leading to the onset of intermittent chaos. Thus, while both contribute significantly to the understanding of chaotic features in QS, the latter offers a more direct mechanistic link between a specific biological timescale and the appearance of intermittent chaos.

In the present work, we further explore this phenomenon by revealing its emergence in QS-inspired systems with heterogeneous temporal delays. This represents a more complex scenario of temporal heterogeneity, demonstrating how the interplay of multiple, distinct delays can lead to intermittent chaos in QS-inspired systems. Our analysis therefore expands the understanding of intermittent chaos in QS dynamics, moving beyond single-delay mechanisms to consider more biologically realistic scenarios with varied response times.

These studies collectively underscore that intermittent chaos remains a less explored area within QS dynamics, representing a significant avenue for future research. This suggests that intermittent chaos may play a crucial role in the flexibility and adaptability of bacterial communication, particularly in fluctuating environments, and warrants further in-depth investigation.

Our findings suggest that identifying specific ``chaotic windows'' within the parameter space of varying delay values can be quite challenging. Even small changes in either $\tau_1$ or $\tau_2$ can lead to relatively abrupt transitions between stable, oscillatory, and chaotic regimes, highlighting the system's sensitivity to its initial conditions and the intricate interplay between the delays. These findings may provide useful insights into microbial communication and could potentially inform the development of innovative strategies for targeted manipulation of QS systems. Moreover, the insights gained from this study may prove helpful in the design and engineering of synthetic biological systems where precise control of gene expression and population dynamics is crucial.

While this investigation has provided key insights into the temporal dynamics of QS, the inherent spatial heterogeneity arising from transport phenomena represents an essential aspect of system behaviour. Future research could explore extending the current model to incorporate such spatial heterogeneity within complex interaction topologies. Although these are important factors beyond the scope of this current analysis, their integration is essential for achieving a more comprehensive understanding, thus capturing the spatio-temporal complexity observed in QS systems.}

\section*{Acknowledgements}  {The authors appreciate the reviewers' valuable comments and insightful suggestions.} AFP thanks  the financial support by `Programa CAPSEM I+DT', Faculty of Engineering, UNAM. MAGO would like to thank Colegio de Ciencia y Tecnología for its financial support through project UACM CCYT2023-IMP-05 and CONAHCyT-SNII. VFBM thanks the financial support by Asociación Mexicana de Cultura~A.C.

\bibliographystyle{siamplain} 
\bibliography{references2}

\begin{thebibliography}{10}

\bibitem{paguirre}
{\sc P.~Aguirre, B.~Krauskopf, and H.~Osinga}, {\em Global invariant manifolds
  near a shilnikov homoclinic bifurcation}, Journal of Computational Dynamics,
  1 (2014), pp.~1--38.

\bibitem{arbi2022a}
{\sc A.~Arbi}, {\em {Controllability of delayed discret Fornasini--Marchesini
  model via quantization and random packet dropouts}}, Math. Model. Nat.
  Phenom., 17 (2022).

\bibitem{arbi2022b}
{\sc A.~Arbi}, {\em Dynamics on time scales of wave solutions for nonlinear
  neural networks}, Waves in Random and Complex Media,  (2022), pp.~1--17.

\bibitem{barbarossa}
{\sc M.~Barbarossa, C.~Kutter, A.~Fekete, and M.~Rothballer}, {\em {A delay
  model for quorum sensing of {\em Pseudomonas putida}}}, BioSystems, 102
  (2020), pp.~148--156.

\bibitem{Calleja2017ResonanceDelays}
{\sc R.~C. Calleja, A.~R. Humphries, and B.~Krauskopf}, {\em {Resonance
  phenomena in a scalar delay differential equation with two state-dependent
  delays}}, SIAM Journal on Applied Dynamical Systems, 16 (2017),
  pp.~1474--1513.

\bibitem{Chen2020PeriodicDelay}
{\sc M.~Chen, J.~Ji, H.~Liu, and F.~Yan}, {\em {Periodic Oscillations in the
  Quorum-Sensing System with Time Delay}}, International Journal of Bifurcation
  and Chaos, 30 (2020).

\bibitem{Chen2019}
{\sc M.~Chen, H.~Liu, and F.~Yan}, {\em Oscillatory dynamics mechanism induced
  by protein synthesis time delay in quorum-sensing system}, Physical Review E,
  99 (2019).

\bibitem{chen}
{\sc M.~Chen, H.~Liu, and F.~Yan}, {\em Modelling and analysing biological
  oscillations in quorum sensing networks}, IET Syst. Biol., 14 (2020),
  pp.~190--199.

\bibitem{cicek}
{\sc O.~Cicek, Y.~B. Ozcelik, and A.~Altan}, {\em {A New Approach Based on
  Metaheuristic Optimization Using Chaotic Functional Connectivity Matrices and
  Fractal Dimension Analysis for AI-Driven Detection of Orthodontic Growth and
  Development Stage}}, Fractal Fract, 9 (2025), p.~148.

\bibitem{demarkus2000}
{\sc A.~S. De~Markus}, {\em Detection of the onset of numerical chaotic
  instabilities by lyapunov exponents}, Discrete Dynamics in Nature and
  Society, 6 (2000), pp.~121--128.

\bibitem{elaskar}
{\sc S.~Elaskar and E.~del R\'io}, {\em {Review of Chaotic Intermittency}},
  Symmetry, 15 (2023), p.~1195.

\bibitem{elowitz}
{\sc M.~Elowitz and S.~Leibler}, {\em A synchronized oscillatory network
  oftranscriptional regulator}, Nature, 403 (2000), pp.~335--338.

\bibitem{engelborghs2002numerical}
{\sc K.~Engelborghs, T.~Luzyanina, and D.~Roose}, {\em Numerical bifurcation
  analysis of delay differential equations using dde-biftool}, ACM Transactions
  on Mathematical Software (TOMS), 28 (2002), pp.~1--21.

\bibitem{fergola2006allelopathic}
{\sc P.~Fergola, M.~Cerasuolo, and E.~Beretta}, {\em An allelopathic
  competition model with quorum sensing and delayed toxicant production},
  {Mathematical Biosciences \& Engineering}, 3 (2006), p.~37.

\bibitem{fergola2008influence}
{\sc P.~Fergola, J.~Zhang, M.~Cerasuolo, and Z.~Ma}, {\em On the influence of
  quorum sensing in the competition between bacteria and immune system of
  invertebrates}, in AIP Conference Proceedings, vol.~1028, American Institute
  of Physics, 2008, pp.~215--232.

\bibitem{frederick}
{\sc M.~R. Frederick, C.~Kuttler, B.~A. Hense, and H.~J. Eberl}, {\em A
  mathematical model of quorum sensing regulated eps production in biofilm
  communities}, Theoretical Biology and Medical Modelling, 8 (2011).

\bibitem{ojalvo}
{\sc J.~Garc\'ia-Ojalvo, M.~B. Elowitz, and S.~H. Strogatz}, {\em Modeling a
  syntheticmulticellular clock: repressilators coupled by quorum sensing},
  Proc. Natl. Acad. Sci. USA, 30 (2004), pp.~10955--10960.

\bibitem{gonchenko97}
{\sc S.~V. Gonchenko, D.~V. Turaev, P.~Gaspard, and G.~Nicolis}, {\em
  Complexity in the bifurcation structure of homoclinic loops to a
  saddle-focus}, Nonlinearity, 10 (1997), p.~409.

\bibitem{chofski2016}
{\sc T.~E. Gorochowski}, {\em Agent-based modelling in synthetic biology},
  Essays Biochem., 60 (2016), pp.~325--336.

\bibitem{goryachev}
{\sc A.~Goryachev}, {\em Design principles of the bacterial quorum sensing gene
  networks}, Wiley Interdiscip Rev Syst Biol Med., 1 (2009), pp.~45--60.

\bibitem{grantham1993}
{\sc W.~Grantham and B.~Lee}, {\em A chaotic limit cycle paradox}, Dynamics and
  Control, 3 (1993), pp.~157--171.

\bibitem{gu2005stability}
{\sc K.~Gu, S.-I. Niculescu, and J.~Chen}, {\em On stability crossing curves
  for general systems with two delays}, Journal of mathematical analysis and
  applications, 311 (2005), pp.~231--253.

\bibitem{guevara1983chaos}
{\sc M.~R. Guevara, L.~Glass, M.~C. Mackey, and A.~Shrier}, {\em Chaos in
  neurobiology}, IEEE Transactions on Systems, Man, and Cybernetics, SMC-13
  (1983), pp.~790--798.

\bibitem{hammer2003quorum}
{\sc B.~K. Hammer and B.~L. Bassler}, {\em Quorum sensing controls biofilm
  formation in vibrio cholerae}, Molecular microbiology, 50 (2003),
  pp.~101--104.

\bibitem{hellen2018}
{\sc E.~Hellen and E.~Volkov}, {\em How to couple identical ring oscillators to
  get quasiperiodicity, extended chaos, multistability, and the loss of
  symmetry}, Commun. Nonlinear Sci. Numer. Simulat., 62 (2018), pp.~462--479.

\bibitem{karkaria2022}
{\sc B.~Karkaria, A.~Manhart, A.~Fedorec, and C.~Barnes}, {\em Chaos in
  synthetic microbial communities}, PLOS Computational Biology, 18 (2022),
  p.~e1010548.

\bibitem{lakshmanan2011dynamics}
{\sc M.~Lakshmanan and D.~V. Senthilkumar}, {\em Dynamics of nonlinear
  time-delay systems}, Springer Science \& Business Media, 2011.

\bibitem{lee2013cell}
{\sc J.~Lee, J.~Wu, Y.~Deng, J.~Wang, C.~Wang, J.~Wang, C.~Chang, Y.~Dong,
  P.~Williams, and L.-H. Zhang}, {\em A cell-cell communication signal
  integrates quorum sensing and stress response}, Nature chemical biology, 9
  (2013), p.~339.

\bibitem{lewis2003autoinhibition}
{\sc J.~Lewis}, {\em Autoinhibition with transcriptional delay: a simple
  mechanism for the zebrafish somitogenesis oscillator}, Current Biology, 13
  (2003), pp.~1398--1408.

\bibitem{lyon2004peptide}
{\sc G.~J. Lyon and R.~P. Novick}, {\em Peptide signaling in staphylococcus
  aureus and other gram-positive bacteria}, Peptides, 25 (2004),
  pp.~1389--1403.

\bibitem{harrisvf}
{\sc {M.~Harris and V.~Rivera--Estay and P.~Aguirre and
  V.F.~{Bre\~na}--Medina}}, {\em {Multiple Local and Global Bifurcations and
  Their Role in Quorum Sensing Dynamics}}, The ANZIAM Journal, 67:e17 (2025).

\bibitem{mackey1977oscillation}
{\sc M.~C. Mackey and L.~Glass}, {\em Oscillation and chaos in physiological
  control systems}, Science, 197 (1977), pp.~287--289.

\bibitem{manneville1979intermittency}
{\sc P.~Manneville and Y.~Pomeau}, {\em Intermittency and the lorenz model},
  Physics Letters A, 75 (1979), pp.~1--2.

\bibitem{matlab}
{\sc {MathWorks}}, {\em Matlab}.
\newblock https://la.mathworks.com/products/matlab.html, 2025.

\bibitem{may1974}
{\sc R.~May}, {\em {Biological populations with nonoverlapping generations:
  Stable points, stable cycles, and chaos}}, Science, 17 (1974), pp.~645--647.

\bibitem{michiels2007stability}
{\sc W.~Michiels and S.-I. Niculescu}, {\em Stability and stabilization of
  time-delay systems: an eigenvalue-based approach}, SIAM, 2007.

\bibitem{miller2001quorum}
{\sc M.~B. Miller and B.~L. Bassler}, {\em Quorum sensing in bacteria}, Annual
  Reviews in Microbiology, 55 (2001), pp.~165--199.

\bibitem{monk2003oscillatory}
{\sc N.~A. Monk}, {\em Oscillatory expression of hes1, p53, and nf-$\kappa$b
  driven by transcriptional time delays}, Current Biology, 13 (2003),
  pp.~1409--1413.

\bibitem{moura2019}
{\sc P.~Moura-Alves, A.~Puyskens, A.~Stinn, M.~Klemm, U.~Guhlich-Bornhof,
  A.~Dorhoi, J.~Furkert, A.~Kreuchwig, J.~Protze, L.~Lozza, G.~Pei, P.~Saikali,
  C.~Perdomo, H.~Mollenkopf, R.~Hurwitz, F.~Kirschhoefer, G.~Brenner-Weiss,
  J.~Weiner~3rd, H.~Oschkinat, M.~Kolbe, G.~Krause, and S.~Kaufmann}, {\em
  {Host monitoring of quorum sensing during {\em Pseudomonas aeruginosa}
  infection}}, Science, 366 (2019).

\bibitem{Mukherjee}
{\sc M.~Mukherjee and B.~L. Bassler}, {\em Bacterial quorum sensing in complex
  and dynamically changing environments}, Nat Rev Microbiol., 17 (2019),
  pp.~371--382.

\bibitem{nealson1970cellular}
{\sc K.~H. Nealson, T.~Platt, and J.~W. Hastings}, {\em Cellular control of the
  synthesis and activity of the bacterial luminescent system}, Journal of
  bacteriology, 104 (1970), pp.~313--322.

\bibitem{newhouse1978occurrence}
{\sc S.~Newhouse, D.~Ruelle, and F.~Takens}, {\em {Occurrence of strange AxiomA
  attractors near quasi periodic flows on $T^m$, $m\geq3$}}, Communications in
  Mathematical Physics, 64 (1978), pp.~35--40.

\bibitem{nguimdo}
{\sc R.~Nguimdo}, {\em {Constructing Hopf bifurcations lines for the stability
  of onlineal systems with two time delays}}, Phys. Rev. E, 97 (2018),
  pp.~1--7.

\bibitem{otto2008}
{\sc M.~Otto}, {\em Staphylococcal biofilms}, vol.~322, Springer, Berlin,
  Heidelberg, 2008, ch.~{In: Romeo, T., Ed., Bacterial Biofilms. Current Topics
  in Microbiology and Immunology}, pp.~207--228.

\bibitem{ozcelik}
{\sc Y.~B. Ozcelik and A.~Altan}, {\em {Overcoming Nonlinear Dynamics in
  Diabetic Retinopathy Classification: A Robust AI-Based Model with Chaotic
  Swarm Intelligence Optimization and Recurrent Long Short-Term Memory}},
  Fractal Fract., 7 (2023), p.~598.

\bibitem{parsek2003}
{\sc M.~Parsek and P.~Singh}, {\em Bacterial biofilms: an emerging link to
  disease pathogenesis.}, Annu Rev Microbiol., 57 (2003), pp.~677--701.

\bibitem{judithpv}
{\sc J.~Perez-Velazquez, B.~Quinones, B.~A. Hense, and C.~Kuttler}, {\em A
  mathematical model to investigate quorum sensing regulation and its
  heterogeneity in pseudomonas syringae on leaves}, Ecological Complexity, 21
  (2015), pp.~128--141.

\bibitem{rosenstein1993practical}
{\sc M.~T. Rosenstein, J.~J. Collins, and C.~J. De~Luca}, {\em A practical
  method for calculating largest lyapunov exponents from small data sets},
  Physica D: Nonlinear Phenomena, 65 (1993), pp.~117--134.

\bibitem{singh2015}
{\sc H.~Singh and P.~Parmananda}, {\em Crowd synchrony in chaotic oscillators},
  Nonlinear Dyn., 80 (2015), pp.~767--776.

\bibitem{hyperchaos}
{\sc N.~Stankevich and E.~Volkov}, {\em Chaos-hyperchaos transition in three
  identical quorum-sensing mean-field coupled ring oscillators}, Chaos, 31
  (2021), p.~103112.

\bibitem{stefanski2010}
{\sc K.~Stefanski, K.~Buszko, and K.~Piecyk}, {\em Transient chaos measurements
  using finite-time lyapunov exponents}, Chaos, 20 (2010), p.~033117.

\bibitem{suzuki2016periodic}
{\sc Y.~Suzuki, M.~Lu, E.~Ben-Jacob, and J.~N. Onuchic}, {\em Periodic,
  quasi-periodic and chaotic dynamics in simple gene elements with time
  delays}, Scientific reports, 6 (2016), p.~21037.

\bibitem{tamsir2011}
{\sc A.~Tamsir, J.~Tabor, and C.~Voight}, {\em Robust multicellular computing
  using genetically encoded nor gates and chemical 'wires'}, Nature, 469
  (2011), pp.~212--215.

\bibitem{walpole2012probabilidad}
{\sc R.~Walpole, R.~Myers, and S.~Myers}, {\em Probabilidad y estad{\'\i}stica
  para ingenier{\'\i}a y ciencias}, Pearson educaci{\'o}n, 2012.

\bibitem{Wernecke_2019}
{\sc H.~Wernecke, B.~S{\'a}ndor, and C.~Gros}, {\em Chaos in time delay
  systems, an educational review}, Physics Reports, 824 (2019), pp.~1--40,
  \url{http://dx.doi.org/10.1016/j.physrep.2019.08.001}.

\bibitem{seb05}
{\sc S.~Wieczorek and B.~Krauskopf}, {\em Bifurcations of n-homoclinic orbits
  in optically injected lasers}, Nonlinearity, 18 (2005), p.~1095.

\bibitem{yanchuk2009delay}
{\sc S.~Yanchuk and P.~Perlikowski}, {\em Delay and periodicity}, Physical
  review E, 79 (2009), p.~046221.

\bibitem{you2004}
{\sc {You, L. and Cox, R.S. 3rd, and Weiss, R. and Arnold, F.H.}}, {\em
  Programmed population control by cell-cell communication and regulated
  killing}, Nature, 428 (2004), pp.~868--871.

\bibitem{zakharova2022}
{\sc A.~Zakharova, G.~Strelkova, E.~Sch{\"o}ll, and J.~Kurths}, {\em
  {Introduction to focus issue: In memory of Vadim S. Anishchenko: Statistical
  physics and nonlinear dynamics of complex systems}}, Chaos, 32 (2022),
  p.~010401.

\bibitem{zhang2017oscillatory}
{\sc Y.~Zhang, H.~Liu, F.~Yan, and J.~Zhou}, {\em Oscillatory dynamics of p38
  activity with transcriptional and translational time delays}, Scientific
  reports, 7 (2017), pp.~1--11.

\bibitem{zhang2016oscillatory}
{\sc Y.~Zhang, H.~Liu, and J.~Zhou}, {\em Oscillatory expression in escherichia
  coli mediated by micrornas with transcriptional and translational time
  delays}, IET Systems Biology, 10 (2016), pp.~203--209.

\bibitem{Zhang20082834}
{\sc Z.~Zhang, Y.~Suo, J.~Peng, and J.~Zhang}, {\em Analysis of a periodic
  bacteria-immunity model with delayed quorum sensing}, Computers and
  Mathematics with Applications, 56 (2008), pp.~2834--2847.

\bibitem{Zhonghua2009201}
{\sc Z.~Zhonghua, S.~Yaohong, Z.~Juan, and P.~Jigeng}, {\em A bacteria-immunity
  system with delayed quorum sensing}, Journal of Applied Mathematics and
  Computing, 30 (2009), pp.~201--217.

\end{thebibliography}

\end{document}